\documentclass[12pt,preprint]{aastex}
\newcommand{\lsim}         {\mbox{$_<\atop^{\sim}$}}
\slugcomment{ApJ, in press, May 22, 2007}
\begin{document}

\title{Determination of the Far-Infrared Cosmic Background
Using {\it COBE}/DIRBE and WHAM Data} 

\author{
N. Odegard,\altaffilmark{1}
R. G. Arendt,\altaffilmark{2}
E. Dwek,\altaffilmark{3}
L. M. Haffner,\altaffilmark{4}
M. G. Hauser,\altaffilmark{5}
R. J. Reynolds\altaffilmark{4}
}
\altaffiltext{1}{ADNET Systems Inc., Code 665, NASA GSFC, Greenbelt, MD 20771; nils.odegard@gsfc.nasa.gov.}
\altaffiltext{2}{SSAI, Code 665, NASA GSFC, Greenbelt, MD 20771.} 
\altaffiltext{3}{Code 665, NASA GSFC, Greenbelt, MD 20771.}
\altaffiltext{4}{University of Wisconsin, Astronomy Department, Madison, WI 53706.}
\altaffiltext{5}{Space Telescope Science Institute, 3700 San Martin Drive, 
Baltimore MD, 21218.}

\begin{abstract}
Determination of the cosmic infrared background (CIB) at far infrared 
wavelengths using {\it COBE}/DIRBE data is limited by 
the accuracy to which foreground interplanetary and Galactic dust 
emission can be modeled and subtracted. Previous determinations 
of the far infrared CIB (e.g., Hauser \break et al. 1998) were based on the 
detection of residual isotropic emission in skymaps from which the 
emission from interplanetary dust and the neutral interstellar 
medium were removed.  In this paper we use the
Wisconsin H$\alpha$ Mapper (WHAM) Northern Sky Survey as a 
tracer of the ionized medium to examine the effect of this foreground 
component on determination of the CIB. We decompose the DIRBE 
far infrared data for five high Galactic latitude regions into H~I-- 
and H$\alpha$--correlated components and a residual component.
Based on {\it FUSE} H$_2$ absorption line observations, the contribution of
an H$_2$--correlated component is expected to be negligible.
We find the H$\alpha$--correlated component to be consistent with zero
for each region, and we find that addition of an H$\alpha$--
correlated component in modeling the foreground emission has negligible 
effect on derived CIB results.  
Our CIB detections and 2$\sigma$ upper limits are essentially the same as 
those derived by Hauser et al. and are
given by $\nu I_{\nu}($nW m$^{-2}$ sr$^{-1}$)  $<$ 75, 
$<$ 32, 
25$\pm$8, and 13$\pm$3 at $\lambda$ = 60, 100, 140, and 240~$\mu$m, respectively.
Our residuals have not been subjected to a detailed anisotropy test, so our CIB 
results do not supersede those of Hauser et al.
We derive upper limits on the 100~$\mu$m emissivity 
of the ionized medium that are typically about 40\% of the 100~$\mu$m
emissivity of the neutral atomic medium.  This low value may be caused in part by
a lower dust-to-gas mass ratio in the ionized medium than in the neutral medium,
and in part by a shortcoming of using H$\alpha$ intensity as a tracer of far infrared 
emission.  If H$\alpha$ is not a reliable tracer, 
our analysis would underestimate the emissivity of the ionized medium, and both
our analysis and the Hauser et al. analysis may slightly overestimate the CIB. 
We estimate the possible effect for the CIB to be only about 5\%, which is much 
smaller than the quoted uncertainties.  From a comparison of the Hauser et al. 
CIB results with the integrated galaxy brightness from {\it Spitzer} source counts,
we obtain 2$\sigma$ upper limits on a possible diffuse CIB component that 
are 26 nW m$^{-2}$ sr$^{-1}$ at 140 $\micron$ and 8.5~nW~m$^{-2}$~sr$^{-1}$ at 
240 $\micron$.

\end{abstract}


\section{Introduction}
The diffuse cosmic infrared background (CIB) consists of the cumulative 
energy releases in the universe that have either been redshifted, or
absorbed and reradiated by dust, into the infrared (IR) wavelength
region.  The CIB therefore provides important constraints on the rates of 
nuclear and gravitational energy release, as well as more exotic forms
of energy release, over the history of the universe. 
Over the past few years, analyses of data obtained with the Diffuse Infrared 
Background Experiment (DIRBE) and the Far Infrared Absolute Spectrophotometer 
(FIRAS) onboard the {\it Cosmic Background
Explorer} ({\it COBE}) satellite have provided the first measurements of the 
cosmic background in the far infrared to submillimeter wavelength region 
(Puget et al. 1996, Schlegel et al. 1998, Fixsen et al. 1998, Hauser et al. 
1998, Lagache et al. 1999, Lagache et al. 2000). A detailed description of 
the {\it COBE} instruments 
and the {\it COBE} mission is given by Boggess et al. (1992), Silverberg (1993), 
Mather, Fixsen, \& Shafer (1993), Hauser et al. (1997), and Brodd et al. (1997). 
Recent reviews covering the history of the quest for the CIB, the current 
detections and limits on its spectrum, and the astrophysical implications are
given by Hauser and Dwek (2001), Kashlinsky (2005), and Lagache, Puget, and
Dole (2005).
 
The far infrared CIB measurements are limited by the accuracy to which
foreground interplanetary and Galactic emission can be modeled and
subtracted from the {\it COBE} data.  Emission from interplanetary dust is the 
dominant foreground below about 100 $\micron$ and emission from interstellar
dust is the dominant one at longer wavelengths.  
 
Different models with different degrees of complexity have been used to remove 
the interplanetary dust (IPD) emission from the {\it COBE} maps. Puget et al. (1996) 
and Schlegel et al. (1998) relied only on the spatial characteristics of the 
IPD emission and subtracted a scaled template based on the DIRBE 
25 $\mu$m sky map. The detections reported by Hauser et al. (1998) and Fixsen 
et al. (1998) used the IPD model of Kelsall et al. (1998), which was fit to the 
time variation of the DIRBE data caused by the motion of the earth through the 
interplanetary dust cloud. The uncertainty in the zero level of the emission
predicted by this model makes a major contribution to the uncertainty of the
related CIB measurements.  Wright (1998) and Gorjian, Wright, and Chary (2000) 
modeled the IPD using a method similar to that of Kelsall et al., with an added 
constraint that the residual 25 $\micron$ intensity after zodiacal light subtraction 
be zero at high Galactic latitudes.

Removal of the emission from interstellar dust requires an interstellar 
medium (ISM) template that has a well-defined zero level and correlates well 
with the spatial variation of IR emission in the {\it COBE} maps.  
Before the completion of the Wisconsin H-Alpha Mapper (WHAM) survey, 
observations of Galactic H~I emission provided the best ISM template for 
this purpose (Puget et al. 1996; Schlegel et al. 1998; and Hauser et al. 1998). 
For example, Hauser et al. (1998) determined correlations of IR intensity with 
Galactic H~I column density for selected regions at high Galactic latitude and 
high ecliptic latitude, using data in the DIRBE 100 $\micron$, 140 $\micron$, 
and 240 $\micron$ bands.
For each region and each wavelength band, the H~I--correlated component of the
infrared emission was subtracted from the data.  Careful error analysis,
including estimates of systematic error in subtraction of the interplanetary
and Galactic foregrounds, showed that the mean residual intensity was 
significantly (more than 3$\sigma$) greater than zero at all three wavelengths.
The residual intensity passed tests for isotropy at 140 $\micron$ and 240 
$\micron$, so detection of the CIB was claimed at these wavelengths.

  This method of subtracting Galactic foreground emission is subject to error
if the ratio of Galactic foreground emission to Galactic H~I column density
varies over the region studied.  Such variation could occur if there is
emission from dust associated with molecular or ionized gas, and this 
emission is not entirely correlated with H~I column density.  Based on the 
FUSE H$_2$ absorption line study of Gillmon et al. (2006) and Gillmon and Shull 
(2006), H$_2$ column density is expected to be negligible compared to H~I 
column density over a large 
fraction of the high latitude sky, and to account for 1\% to 30\% of total 
H column density for cirrus features brighter than $1.5-3$ MJy sr$^{-1}$ in the 
temperature-corrected \break 100 $\micron$ map of Schlegel, Finkbeiner, and
Davis (1998).  Most far infrared CIB 
determinations have used restrictions on Galactic latitude, H~I column 
density, and/or far infrared color to exclude lines of sight that may 
contain significant emission from dust associated with molecular gas.
On the other hand, the warm ionized medium (WIM) is known to be prevalent 
at high Galactic latitudes.  Available data indicates that H~II column 
density is on average about one-third of H~I column density (Reynolds 1991a),
and correlation studies using H$\alpha$ as a tracer suggest that a significant
fraction of the H~II is not spatially correlated with H~I (e.g., Reynolds et al. 
1995, Arendt et al. 1998, Lagache et al. 2000). 
The ionized gas is 
expected to be subject to approximately the same interstellar radiation 
field as the neutral gas, and the depletion study of Howk and Savage 
(1999) shows that the dust--to--gas mass ratio in the WIM may be similar to 
that in the warm neutral medium.
Thus, the infrared emissivity per H nucleus may be similar in the ionized and
diffuse H~I phases of the ISM, and infrared emission from the ionized phase 
may have a significant effect on any CIB determination that is solely based on 
correlation with an H~I template.
 
The paucity of tracers of the ionized gas at high latitudes precluded any 
definitive measurement of the amount of IR emission from dust in this gas 
phase.  Nevertheless, several methods were used to estimate and subtract its
contribution to the foreground ISM emission: (1) Puget et al. (1996) and 
Lagache et al. (1999) identified the H~II emission component with a spatially 
varying component obtained after subtraction of an H~I correlated 
foreground component from {\it COBE}/FIRAS data. This residual component exhibited a 
csc($\vert b \vert $) 
dependence (Boulanger et al. 1996), consistent with that expected from a plane 
parallel layer of ionized gas. (2) Fixsen et al. (1998) used maps of
H~I column density and [C II] 158~$\micron$ line emission as templates
to model the IR emission from the neutral and ionized gas phases. They found
that essentially none of the high latitude emission observed by FIRAS 
correlated with the [C II] template, and their final CIB spectrum was 
more than two times greater than the H~II emission subtracted CIB spectrum 
of Puget et al. (1996) at $\lambda < 240 \micron$.
(3) Arendt et al. (1998) derived an upper limit to the 100 $\micron$ emissivity 
per H nucleus in the ionized medium from a correlation analysis of H$\alpha$, \break
H~I, and DIRBE maps for a $10\arcdeg \times 12\arcdeg$ region centered at 
$l = 144\arcdeg$, $b = -21\arcdeg$.  The derived $3\sigma$ upper limit was equal 
to 3/4 of the 100 $\micron$ emissivity per H nucleus for the neutral atomic gas 
in the same region.  Assuming this limit is valid for the Lockman hole region, 
they used available H$\alpha$ and pulsar dispersion measure data for the 
Lockman hole to place
an upper limit of 4 nW m$^{-2}$ sr$^{-1}$ on the possible contribution of the ionized
gas phase to the 100 $\micron$ foreground emission.  This was scaled to obtain 
upper limits of 5 and 2 nW m$^{-2}$ sr$^{-1}$ at \break 140 and 240 $\micron$, respectively,
assuming the spectrum of emission from the ionized phase has the same shape as
that of the neutral atomic phase.  These upper limits are comparable to the
overall uncertainties in the CIB determinations.  They were noted as possible
errors by Hauser et al. (1998) but were not included in their quoted CIB
uncertainties.  If they had been included, their reported 240 $\mu$m CIB
value would still be a 3$\sigma$ detection.
 
The Wisconsin H-Alpha Mapper (WHAM) northern sky survey provided the first 
H$\alpha$ map sensitive enough to trace the ionized gas phase of the ISM
at high Galactic latitude (Haffner et al. 2003).  
Lagache et al. (2000) used preliminary WHAM survey data
to decompose the 100 to 1000 $\micron$ DIRBE and FIRAS data at high latitude
into H$\alpha$-correlated, H~I-correlated, and isotropic components.
The regions studied cover about 2\% of the sky in the range
$25\arcdeg < \vert b \vert < 50\arcdeg$.  They found a significant
H$\alpha$-correlated component with dust temperature of 17.2 K,
very similar to that of the H~I-correlated component.  Assuming a
constant electron density of 0.08 cm$^{-3}$ and an electron temperature of 
8000~K for the ionized gas, they derived an infrared emissivity per H~nucleus
for this phase that is similar to that of the neutral phase.  The CIB 
spectrum they determined from analysis of the FIRAS data is consistent 
with that of Fixsen et al. (1998).  The mean residual intensities
they found at 100, 140, and 240 $\micron$ from analysis of the DIRBE data 
are consistent with the results of Hauser et al. (1998), although their 
uncertainties are larger.

  The agreement between these latest CIB determinations is encouraging,
but the disagreement between the Arendt et al. (1998) and Lagache et al. 
(2000) results for the \break 100 $\micron$ emissivity of the ionized medium is 
a matter of concern.  In this paper, we address the possible effects of
the ionized medium on the Hauser et al. (1998) CIB results by including 
an H$\alpha$-correlated foreground component in an analysis that
is otherwise similar to the DIRBE team analysis.  We make use of 
H$\alpha$ data from the WHAM Northern Sky Survey.  The
paper is organized as follows.  In \S2 we describe the data sets and sky 
areas used in the analysis.  The decomposition method is described in \S3.
From the decompositions we derive the emissivity per H nucleus for the H~I 
and H~II phases of the ISM as well as the residual emission for the 
different DIRBE bands.  In \S4 we present these results and compare them with
results of previous studies.  In \S5 we discuss the emissivity results and we 
discuss possible systematic errors if H$\alpha$ is not
a reliable tracer of far infrared emission.
Our results and conclusions are summarized in \S6.

\section{Data Sets}

The regions of the sky analyzed in this paper are shown in Figure 1.
Three of them were previously analyzed
by Hauser et al. (1998) and Arendt et al. (1998): the Lockman Hole (LH), a 
300 square degree region around the position of lowest H~I column density 
at $l=152\arcdeg, b=+52\arcdeg$; an $8\arcdeg \times 9\arcdeg$ region centered 
on the north ecliptic pole (NEP) at $l=96\arcdeg, b=+30\arcdeg$; and the DIRBE 
high quality B north (HQBN) region at $b > +60\arcdeg$ and $\beta > +45\arcdeg$.
These regions were originally selected because they were expected to have
relatively weak Galactic and interplanetary dust foregrounds, and because 
good quality H~I observations were available for them.  The DIRBE high quality
B south region is not included in our analysis because it is below the 
declination limit of the WHAM Northern Sky Survey.
In addition, we analyze the second quadrant region previously studied by 
Lagache et al. (2000) and a new region in the first quadrant at 
$30\arcdeg < l < 80\arcdeg, 30\arcdeg < b < 41\arcdeg$.  We refer to these
as the Q2 region and Q1 region, respectively.  The Q2 region was originally
selected because preliminary WHAM H$\alpha$ data were available for it.
The Q1 region was selected because it is comparable to the Q2 region in
Galactic latitude, it is at ecliptic latitude greater than 30 degrees,
it has no molecular clouds detected in the CO survey of Hartmann, Magnani,
and Thaddeus 
(1998), and it has no cold infrared excess features that are characteristic 
of molecular clouds in the 100 $\micron$ infrared excess map of Reach et al. 
(1998). 

The datasets used in our analysis are listed in Table 1.  We use DIRBE
60, 100, 140, and 240 $\micron$ mission-averaged skymaps from which the 
interplanetary foreground emission has been subtracted using the model of 
Kelsall et al. (1998).  Foreground Galactic stellar emission is negligible 
at these wavelengths (Hauser et al. 1998, Arendt et al. 1998) and has not
been subtracted from the data.

We use H~I 21-cm line data integrated over a velocity
range that includes all significant Galactic emission, converted to H~I column 
density $N$(H~I) assuming that the line emission is optically thin.  The H~I
data for the LH and NEP regions are from Snowden et al. (1994) and
Elvis, Lockman, and Fassnacht (1994), and were corrected for stray radiation
using the AT\&T Bell Laboratories H~I survey (Stark et al. 1992).  Estimated
1$\sigma$ uncertainties in $N$(H~I) for these regions range from $0.5\times10^{19}$ 
cm$^{-2}$ to $1.0\times10^{19}$ cm$^{-2}$.  The H~I data that we use for the 
other 
regions are from the Leiden-Dwingeloo H~I survey (Hartmann and Burton 1997), 
which has also been corrected for stray radiation (Hartmann et al. 1996).  
We adopt a 1$\sigma$ uncertainty of $1.25\times10^{19}$ cm$^{-2}$ for each position 
in these regions. This is the uncertainty of the stray radiation 
correction that was estimated by Hartmann et al. (1996) for a reference
position in the Lockman Hole region.

We use H$\alpha$ total intensity data from the Wisconsin H-Alpha Mapper (WHAM)
Northern Sky Survey (Haffner et al. 2003), which covers the sky north of 
declination $-30\arcdeg$.
The WHAM instrument has a 1 degree diameter field of view, and the survey
was made on a regular Galactic coordinate grid with pointings separated
by $0.98\arcdeg /$cos $b$ in $l$ and $0.85\arcdeg$ in $b$.  The spectrum for each pointing
was integrated over $-80 < v_{LSR} < 80$ km s$^{-1}$ to obtain total H$\alpha$ 
intensity.  For the regions we study, systematic errors associated with removal 
of geocoronal and atmospheric
emission lines from the spectra can be greater than statistical measurement 
uncertainties.  These errors can vary from night to night, and sometimes cause 
$\sim7\arcdeg\times7\arcdeg$ "blocks" of data taken on particular nights to be
noticeably offset in mean intensity relative to their surroundings. 
We applied offset corrections to affected blocks in the HQBN, Lockman Hole, and Q2 
regions to remove discontinuities in H$\alpha$ intensity at the block boundaries.
For the HQBN and Q2 regions, most blocks appeared to be unaffected and these were 
assumed to set the zero level of the data.    
For the HQBN region, an offset of 0.15 Rayleigh was added to the data in the areas 
($102\arcdeg < l < 117\arcdeg, 65\arcdeg < b < 71\arcdeg$), ($60\arcdeg < l < 90\arcdeg, 
65\arcdeg < b < 71\arcdeg$), and ($l > 117\arcdeg, 60\arcdeg < b < 65\arcdeg$).
For the Q2 region, an offset of 0.3 R was added in ($129\arcdeg < l < 136.6\arcdeg,
41\arcdeg < b < 47\arcdeg$) and an offset of \break 0.4 R was added in 
($136\arcdeg < l < 157\arcdeg, 47\arcdeg < b < 51\arcdeg$).
For the Lockman Hole, the offset corrections are somewhat uncertain and subjective
since many blocks appear to be affected, by varying amounts. Offset corrections 
ranged from -0.5 to 0.1 R, and the zero level is determined by the
WHAM observations of Hausen et al. (2002) for two positions in the region.
Our decomposition results for the Lockman Hole are consistent
with those for the other regions, and omitting this region from our analysis
does not change our derived CIB results significantly.
We adopt a 1$\sigma$ uncertainty of 0.06 Rayleigh for the velocity-integrated, 
offset-corrected H$\alpha$ intensity at each WHAM pointing.
This is near the low end of the range of rms H$\alpha$ dispersion measured 
within observing blocks at high Galactic latitudes.  The dispersion tends to 
increase with increasing mean H$\alpha$ intensity or decreasing latitude, 
presumably due to increasing rms dispersion in Galactic emission.
With the adopted uncertainty, the mean H$\alpha$ signal-to-noise ratio is 
4, 8, 18, 18, and 31 for the LH, HQBN, Q1, Q2, and NEP regions, respectively.

The DIRBE data and Leiden H~I data were interpolated to the WHAM pointing positions
for our analysis.  The angular resolution of the Elvis et al. and Snowden et al. 
H~I data is much better that that of the other datasets, so these data were 
averaged over the WHAM field of view at each WHAM pointing position.
Possible contamination by stellar H$\alpha$ absorption was handled either
by excluding positions with stars brighter than $V=6.5$, or by excluding
positions with H$\alpha$ intensity significantly lower than their surroundings.
These two methods were
compared for the LH and NEP regions, and found to give consistent results.
Some positions with a discrete source detected in the DIRBE data
(the planetary nebulae NGC 6543, and galaxies NGC 3079, 3310, 
3556, 3690, and 4102) were excluded from our analysis.  The planetary
nebula NGC 6210 appears as a bright point source in the WHAM data
(Reynolds et al. 2005) and was also excluded.

Maps of the 100 $\micron$ intensity, H~I column density, and H$\alpha$ intensity
are shown for the Q1 region in Figure 2, and correlation plots are shown for 
this region in Figure 3.  These figures show that the correlation between 
$\nu I_{\nu}$(100 $\micron$) and $N$(H~I) is tighter than that between 
$\nu I_{\nu}$(100 $\micron$) 
and $I$(H$\alpha$) or that between $I$(H$\alpha$) and $N$(H~I).  Similar trends are 
seen for the other regions analyzed in this paper, and for the region around 
$l=144\arcdeg, b=-21\arcdeg$ studied previously (Reynolds et al. 1995, Arendt 
et al. 1998).

\section{Analysis}
\subsection{Decomposition of the Infrared Emission}

The method of analysis follows that used by Arendt et al. (1998) and is similar
to that used by Lagache et al. (2000).
For each DIRBE wavelength band (60, 100, 140, and 240 $\micron$), the infrared 
intensity distribution within a given region is decomposed into a component that is 
correlated with H~I column density, a component that is correlated with H$\alpha$ 
intensity, and an isotropic component.  This is done by
making a least squares fit of the form
\begin{equation}
\nu I_{\nu}(\lambda) = A_1 N({\rm H\,I}) + B_1 I(\rm H\alpha) + \it C_1
\end{equation}
where $I_{\nu}(\lambda)$ is the infrared intensity after interplanetary foreground
subtraction and $A_1$, $B_1$, and $C_1$ are fit parameters. $A_1$ is the mean
infrared emissivity per H atom for dust in the neutral atomic gas phase, $B_1$
is a measure of the mean infrared emissivity of dust in the ionized gas phase, 
and the intercept $C_1$ is the mean residual infrared intensity.  For comparison,
a second decomposition is performed in which an H$\alpha$ correlated component 
is not included, by making a fit of the form
\begin{equation}
\nu I_{\nu}(\lambda) = A_2 N({\rm H\, I}) + C_2.
\end{equation}
Comparison of the derived $C_1$ and $C_2$ values gives the error in the
Galactic foreground subtraction if the H$\alpha$ correlated component
is neglected.

Decomposition of an infrared intensity distribution into three components
as in equation (1) will be successful if the spatial distibutions of $N$(H~I) and
$I$(H$\alpha$) differ significantly from each other, and also differ significantly
from an isotropic distribution.  These conditions are met for each of the
regions studied here.  This is illustrated for the Q1 region by Figures 2 and 3.
The method of analysis also
assumes that H$\alpha$ intensity is a good tracer of far infrared emission from
the ionized medium.  
In \S5, we discuss possible errors in our results if this is not the case.  
Extinction of
the H$\alpha$ emission is another potential source of error, but its effect
is negligible for our regions with the data selection criteria described
below.  We made fits to the 100 $\micron$ data for each region with and 
without a correction to $I$(H$\alpha$) for extinction, and differences in 
the results were insignificant.  We made the worst-case assumption of pure
foreground extinction.  With the optical depth at H$\alpha$ calculated as
$\tau = 0.04 [N$(H~I)/$10^{20}$ cm$^{-2}$], the extinction correction was 
at most 1.22.

The fits are made using an iterative procedure that minimizes $\chi^2$
calculated using measurement uncertainties in the independent and dependent
variables (Press et al. 1992).  Uncertainties in the fit parameters are
determined from the 68\% joint confidence region in parameter space,
using the method of Bard (1974).  For each of the regions except for HQBN, 
data at the highest H~I column densities are excluded from the fitting, as 
described by Arendt et al. (1998) for the LH and NEP regions.  The 
$\nu I_{\nu} (100 \thinspace \micron)$ -- {\it N}(H~I) relation deviates from 
linearity in these regions, with excess 100 $\micron$ emission relative to 
{\it N}(H~I) at the highest H~I column densities.  This type of
relation has been found previously for isolated cirrus clouds and for large
regions of the sky at high Galactic latitude, and the excess 100 $\micron$ 
emission has been attributed to emission from dust associated with molecular
gas or to non-negligible optical depth in the 21-cm line
(e.g., Deul and Burton 1993, Reach, Koo, and Heiles 1994, Boulanger et al. 1996).
For the Lockman Hole, it is consistent with detections of CO line emission 
toward some 100 $\micron$ brightness peaks (Heiles, Reach, and Koo 1988;
Stacy et al. 1991; Reach, Koo, and Heiles 1994).  We exclude data above H~I 
column densities where the $\nu I_{\nu}(100 \thinspace  \micron)$ -- {\it N}(H~I) 
relation begins to deviate from linearity (see Figure 8 of Arendt et al. 1998).
The cut is made at $N({\rm H\, I}) = 1.5\times10^{20}$ cm$^{-2}$,
$5.0\times10^{20}$ cm$^{-2}$, $3.0\times10^{20}$ cm$^{-2}$, and
$3.0\times10^{20}$ cm$^{-2}$ for the LH, NEP, Q1, and Q2 regions, 
respectively. 
Measurement uncertainties are much larger for the DIRBE 140 and 240 $\micron$ 
bands than for the 60 and 100 $\micron$ bands, so less stringent 
{\it N}(H~I) limits were adopted for the LH and NEP fits in these bands, 
$2.0\times10^{20}$ cm$^{-2}$ for the LH and $6.0\times10^{20}$ cm$^{-2}$ 
for the NEP.
With these cuts, the area of the sky used in the 100 $\micron$ analysis 
is 420, 170, 35, 180, and 380 square degrees for the HQBN, LH, NEP,
Q1, and Q2 regions, respectively.

\subsection{Limits on Emission from Dust in H$_2$}

 Our analysis does not allow for possible FIR emission from molecular gas that is 
not correlated with H~I.  This emission is expected to be negligible for  
the WHAM pointings that pass the $N$(H~I) cuts. Gillmon et al. (2006) and Gillmon and  
Shull (2006) reported results from a {\it FUSE} survey of H$_2$ absorption lines  
toward 45 AGNs at $\vert b \vert > 20\arcdeg$.  They compared their derived values  
of the molecular fraction, $f_{H_2} = 2N($H$_2)/[N($H~I$)+2N($H$_2)]$, with values of  
temperature-corrected 100 $\micron$ intensity $D^{T}$ from the map of Schlegel,  
Finkbeiner, and Davis (1998).  ($D^{T}$ is proportional to Galactic dust column 
density.)
The transition from low molecular fractions characteristic of optically thin 
clouds to high values characteristic of H$_2$ self-shielded clouds was found to 
occur over the range $1.5 < D^{T} < 3$ \nobreak MJy sr$^{-1}$, \break with $f_{H_2}$ 
varying between $10^{-6}$ and $10^{-1}$ in this $D^{T}$ range, $f_{H_2}$ less than  
$10^{-3}$ at lower $D^{T}$ values, and $f_{H_2}$ between $10^{-2}$ and 0.3 at  
higher $D^{T}$ values. 
Except for the NEP region, most of the WHAM pointings used in our analysis (after the
$N$(H~I) cuts have been applied) have $D^{T}$ 
less than 1.5 MJy sr$^{-1}$, so $f_{H_2}$ is expected to be generally less than $10^{-3}$ 
and the uncertainty in derived CIB results due to neglect of this component
is negligible.  For each region except the NEP, we estimate that dust associated 
with H$_2$ contributes less than 0.04, 0.05, and 0.02 nW m$^{-2}$ sr$^{-1}$ at 100, 140, 
and 240 $\micron$, respectively.  This assumes that $N$(H$_2$) is constant within a region, 
$2 N$(H$_2) < 10^{-3} \thinspace \times$ mean $N$(H~I), and IR emissivity 
per H nucleus in the H$_2$ phase is given by the slope of the IR -- $N$(H~I) relation.
For the NEP, most of the WHAM pointings that are used have $D^{T}$ in the transition range 
from 1.5 to 3.0 MJy sr$^{-1}$, so $f_{H_2}$ values as large as 0.1 are possible.  However, 
we find no evidence for significant IR emission from H$_2$ associated dust.  The mean 
residual infrared intensities $C_1$ from our analysis of the NEP region are consistent 
with the values found for the other regions, and the 100 $\micron$ -- $N$(H~I) relation 
for the NEP is linear with small scatter over the range of $N$(H~I) used in our analysis 
(see figures 7 and 8 of Arendt et al. 1998).  Excluding the NEP region from our analysis 
would not change our derived CIB results significantly.  

\subsection{Test Against Previous Results}
As a check of our analysis method and software, we performed fits of the
form of equation (1) using data previously analyzed by Lagache et al.
(2000) for the Q2 region.
Lagache et al. performed fits of this form for 122 positions in the
region using DIRBE data with interplanetary foreground subtracted, 
Leiden-Dwingeloo H~I data, and preliminary WHAM data, all smoothed to 
the $\sim7\arcdeg$ resolution of the {\it COBE}/FIRAS. They kindly provided 
us with 
the data.  We performed fits as described above, except to be consistent
with the Lagache et al. analysis, only measurement uncertainties for the dependent 
variable (the infrared intensity) were used in the calculation of
$\chi^2$. Table 2 gives a comparison of our results and the Lagache et al.
results.  The two analyses give the same values for the fit parameters, 
but the values for the fit parameter uncertainties do not agree.  Our 
uncertainties for $A_1$ and $B_1$ are larger than those given by Lagache et al.
because our uncertainty calculation allows for uncertainty due to coupling 
between 
the parameters.  (The uncertainty quoted for $C_1$ by Lagache et al. is not a
statistical uncertainty from the fitting, but was determined from the distribution
of residual intensity values, so comparison with our uncertainty value for 
$C_1$ is not meaningful.)  We conclude from this test that differences between 
our results in \S4 and those of Lagache et al. are not the result of software errors.

\section{Results}

  Parameters from our fits for the five regions are listed in Table 3. Positions 
with H I column density greater that the cuts described in section 3.1 were excluded 
from the fitting. For each region and each wavelength, the first row in the table
gives results from the three-component fit (equation 1) and the second row 
gives results from the two-component fit (equation 2).  The uncertainties listed 
are statistical uncertainties from the 68\% joint confidence region in parameter 
space; the $C_1$ and $C_2$ uncertainties do not include systematic uncertainties 
that need to be included in determining
the total uncertainty for a CIB measurement.  Column 7 of the table lists the
number of independent WHAM pointings used for each fit.
Sample correlation plots showing the fits to the 100 $\micron$ data for the 
Q1 and Q2 regions are shown in Figure 4, and parameters from the three-component 
fits for each region are plotted as a function of wavelength in Figures 5 and 6.  

For all regions in all wavelength bands analyzed, we find that the mean 
residual intensity is nearly the same whether an H$\alpha$-correlated
component is included in the fitting or not (the $C_1$ and $C_2$ values are
in close agreement), and the quality of the fit is nearly the same in the 
two cases.  Also, we do not detect significant H$\alpha$-correlated infrared 
emission; the $B_1$ values are consistent with zero within the uncertainties,
and the $A_1$ and $A_2$ values are nearly the same.  The lack of 
H$\alpha$-correlated 100 $\micron$ emission is illustrated for the Q1 region
in Figures 2 and 3.  No correlation is seen between the
H$\alpha$ map in Figure 2c and and the map in Figure 2d of residual 100 $\micron$ 
emission after subtraction of H~I correlated 100 $\micron$ emission.  This is 
also shown by the correlation plot in Figure 3d.

\subsection{Residual Intensities and CIB Measurements}

Figure 7 shows the residual intensity averaged over the five
regions as a function of wavelength for each fit type; the $C_1$ and $C_2$ 
values were each averaged over the regions using weighting by 1/$\sigma^2$.
The weighted-average residual intensities for the different fit types agree 
within 1$\sigma$ statistical uncertainties at all wavelengths, and the agreement 
is within 2 nW m$^{-2}$ sr$^{-1}$ at 60, 100, and 240 $\micron$.  
We conclude that addition of an H$\alpha$-correlated component in modeling the foreground
emission at high Galactic latitude has negligible effect on derived CIB results.

To assess whether the weighted-average values for the residual emission 
can be identified as CIB measurements, we calculated the total uncertainty 
for these values following the method used by Hauser et al. (1998).  The
total uncertainty was calculated as the quadrature sum of the statistical uncertainty, 
interplanetary foreground subtraction uncertainty, and DIRBE detector offset uncertainty.
Magnitudes of the latter two uncertainties were taken from Table 6 of Arendt et al. (1998). 
(DIRBE gain uncertainty is not included because it has the same multiplicative
effect on the mean residual and its total uncertainty, and so does not affect the 
signal-to-noise ratio. This uncertainty is 10--14\% for the DIRBE bands used here.)
Based on these total uncertainties, the mean residual emission is more than
3$\sigma$ greater than zero at 140 $\micron$ and 240 $\micron$ for both the 
two-component fits and the three-component fits, and at 100 $\micron$ for the 
three-component fits.  The residual intensity values for the five different regions 
(Table 3 and Figure 6) are consistent with isotropy at 140~$\micron$ and 240~$\micron$
(the values are compatible within their 1$\sigma$ uncertainties).  We have not performed 
detailed anisotropy tests on the maps of residual intensity from our analysis, 
such as the tests performed by Hauser et al. (1998).  However, based on the reduced 
chi-square values in Table 3, significant anisotropy is present in the Q1 and Q2 
regions at 60 $\micron$ and in all five regions at 100 $\micron$.

Table 4 lists the upper limits and detections of the CIB at 60, 100, 140, and
240 $\micron$ from Hauser et al. (1998), from our
two-component and three-component fits, and from Lagache et al. (2000). For
cases where the mean residual intensity does not exceed zero by greater than
3$\sigma$ or the residual intensity distribution has been shown to be 
anisotropic, the table lists a 2$\sigma$ upper limit on the CIB followed
by the mean residual intensity and its 1$\sigma$ uncertainty in parentheses.  Our 
results are consistent with those of Hauser et al.  Our residuals have not passed
a detailed anisotropy test, and in some cases our quoted uncertainties are slightly 
larger than those of Hauser et al., so our CIB results do not supersede the
Hauser et al. results.

  If H$\alpha$ intensity is not a good tracer of infrared emission from 
dust in the WIM, our three-component fits may not account for emission from
the WIM much better than fits without an H$\alpha$ correlated component do.
Any emission from the WIM that is not correlated with H$\alpha$ or H~I would 
contribute to the residual component, so our results would overestimate the
CIB.  The agreement with the Hauser et al. (1998) results could be 
because their analysis overestimates the CIB by a similar amount.
In \S5.4 we discuss evidence that H$\alpha$ may not be a good tracer,
and we estimate the possible effect of emission from the WIM on the CIB results 
from our analysis and from that of Hauser et al.  For each analysis, we estimate the effect to be 
only about 5\% at 140 and 240 $\micron$, which is much smaller than the quoted uncertainties.

\subsubsection{Systematic Uncertainties}

Although this paper is primarily concerned with the effect of the ionized ISM on CIB determination,
we discuss here other systematic uncertainties involving the level of the H~I cut, choice of 
interplanetary dust model, and choice of photometric calibration.  
Our results for the effect of the ionized ISM have no significant dependence
on any of these uncertainties.

Arendt et al. (1998) found that the sensitivity of the Hauser et al. (1998) CIB results to the H~I cut 
is negligible compared to other sources of uncertainty.  For the Lockman Hole, for example, the 
mean 100 $\micron$ residual intensity was found to vary by less than 0.7~nW~m$^{-2}$~sr$^{-1}$ when 
the cut was varied from 1.0 to $2.0\times10^{20}$ cm$^{-2}$.
We have investigated the sensitivity of the CIB results from our three-component analysis to the 
H~I cut used for our regions that were not included in the Hauser et al. analysis, the Q1 and Q2 
regions. The effect of varying the cut for these regions over the range from 2.5 to 
$3.5\times10^{20}$ cm$^{-2}$ is small.  The derived mean residual changes by +0.1, -0.7, -0.1, and +0.7 
nW m$^{-2}$ sr$^{-1}$ at 60, 100, 140, 240 $\micron$, respectively, when the cut is changed from 
3.0 to $2.5\times10^{20}$ cm$^{-2}$.  These changes are at most 25\% of the uncertainties of the mean 
residuals given in Table 4.  The mean residual changes by -0.8, -0.5, -3.0, and -1.3 
nW m$^{-2}$ sr$^{-1}$ in these bands when the cut is changed from 3.0 to $3.5\times10^{20}$ cm$^{-2}$.
These changes are at most 45\% of the quoted uncertainties of the mean residuals.

The effect of choice of interplanetary dust model is shown in Table 5.  This table compares mean 
residual intensities from Hauser et al. (1998) with those obtained by Wright (2004).
The main difference between these analyses is that Hauser et al. used the interplanetary dust model of 
Kelsall et al. (1998) and Wright used the model of Gorjian et al. (2000).  Each of these models 
was obtained by fitting the time variation observed over the whole sky in each of the DIRBE bands with 
a parameterized model of the dust cloud, but Gorjian et al. added a constraint that the residual 25 $\micron$
intensity after zodiacal light subtraction be zero at high Galactic latitude.  This assumed that the 
25 $\micron$ CIB is negligible, based on CIB upper limits inferred from TeV gamma ray observations of Mrk~501.
Table 5 also lists the uncertainty of the zodiacal light subtraction for the Hauser et al. analysis, as
estimated by Kelsall et al.  This was obtained by comparing results from a series of 
models with different geometries for the density distribution of the dust cloud, which gave 
comparable quality fits to the time variation of the DIRBE data.  The difference between the 
Hauser et al. results and the Wright results is not significant relative to the uncertainty in
zodiacal light subtraction, or relative to the total uncertainty of the mean residual.  

Hauser et al. (1998) noted the effects on CIB results at 140 and 240 $\micron$ if the DIRBE data are
transformed to the FIRAS photometric system.  The CIB values are affected by differences in zero point 
and gain between the DIRBE and FIRAS photometric systems, which are not significant relative 
to the zero point uncertainties and gain uncertainties for the two systems.
Hauser et al. found that transforming the DIRBE CIB results to the FIRAS 
system would reduce the CIB value from 25.0 to 15.0 nW m$^{-2}$ sr$^{-1}$ at 140 $\micron$, and from
13.6 to 12.7 nW m$^{-2}$ sr$^{-1}$ at 240 $\micron$.  
The DIRBE team chose to use the DIRBE photometric system to report its results since comparing the 
two photometric systems has its own uncertainties associated with the need to integrate the DIRBE 
map over the FIRAS beam, and the FIRAS spectrum over the DIRBE spectral response.  
The DIRBE team chose not to introduce this additional uncertainty, and we made the same choice 
for this paper.

\subsubsection{Comparison with Spitzer Source Counts}

Recently, Dole et al. (2006) used {\it Spitzer} MIPS observations to measure the contribution of galaxies
selected at 24 $\micron$ to the CIB at 70 $\micron$ and 160 $\micron$.  They found that sources 
brighter than 60 $\mu$Jy at 24 $\micron$ contribute 5.9 $\pm$ 0.9 and 10.7 $\pm$ 1.6 nW m$^{-2}$ sr$^{-1}$ 
to the CIB at 70 $\micron$ and 160 $\micron$, respectively.  With an extrapolation of the 24 $\micron$ 
source counts below the 60 $\mu$Jy detection threshold,
they estimated that the full population of 24 $\micron$ sources contributes 
7.1 $\pm$ 1.0 and 13.4 $\pm$ 1.7 nW m$^{-2}$ sr$^{-1}$ at 70 and 160 $\micron$.  They noted that these 
should be regarded as lower limits to the CIB since there may be contributions from a diffuse 
background component or from sources missed by the 24 $\micron$ selection.  They estimated that these
contributions account for less than $\sim20\%$ of the far infrared CIB.

A diffuse component of the CIB could be produced by processes such as emission from intergalactic dust or 
radiative decay of primordial particles.  Here we compare the Dole et al. integrated galaxy brightness at 
160 $\micron$ with DIRBE CIB measurements at 140 $\micron$ and 240 $\micron$ to estimate upper limits on 
emission from a diffuse component.  We scale the quoted 
160~$\micron$ intensity of 13.4 $\pm$ 1.7 nW m$^{-2}$ sr$^{-1}$ to 140~$\micron$ and 240~$\micron$ using 
the shape of the 
model spectral energy distribution from figure 13 of Dole et al., which is based on the Lagache et al. 
(2004) galaxy evolution model.  This spectrum is also used to apply color corrections so the intensities 
can be compared with the quoted DIRBE CIB values, which assume a spectral shape of $\nu I_{\nu} =$ constant
over the DIRBE bandpass.  We obtain integrated galaxy intensity 
values of 13.7 $\pm$ 1.7 nW m$^{-2}$ sr$^{-1}$ at 140 $\micron$ and 10.7 $\pm$ 1.4~nW~m$^{-2}$~sr$^{-1}$ at 
240 $\micron$.  The uncertainties here do not include any uncertainty in the shape of the adopted spectral 
energy distribution.  Table 6 compares these values with the DIRBE CIB results, and with DIRBE CIB 
results transformed to the FIRAS photometric system (Hauser et al. 1998).
The table lists values for the fractional contribution of the integrated galaxy brightness to the CIB,
and the difference between the CIB and the integrated galaxy brightness.
Using the CIB results on the DIRBE photometric system yields 2$\sigma$ 
upper limits for a diffuse CIB component of 26 nW m$^{-2}$ sr$^{-1}$ at 140 $\micron$ and 
8.5~nW~m$^{-2}$~sr$^{-1}$ at 240 $\micron$.

\subsection{Infrared Emissivity of the Ionized Medium}

Infrared emissivity results from our three-component fits are shown as a
function of wavelength in Figure 5. The derived values of emissivity per H
atom for the neutral atomic gas phase, $\epsilon$(H$^0$), are comparable to previous 
determinations
from high latitude IR -- H~I correlation studies (e.g., Dwek et al. 1997, Reach 
et al. 1998).  The values of emissivity per H$^{+}$ ion for the ionized phase, 
$\epsilon$(H$^+$), were obtained from the $B_1$ values in Table 3 using a conversion
factor of $I$(H$\alpha$)/$N$(H$^{+})$ = 1.15 Rayleighs/$10^{20}$ cm$^{-2}$.
This is the mean ratio of H$\alpha$ intensity to pulsar dispersion measure
found by Reynolds (1991b) for four high latitude lines of sight toward pulsars 
at $z >$ 4 kpc.  It corresponds to an effective electron density, 
$n_{eff} \equiv \int n_e^2 ds/\int n_e ds$, of 0.08 cm$^{-3}$ for an electron 
temperature of 8000 K and no extinction.  The $I$(H$\alpha$)/$N$(H$^{+})$ ratio 
ranges from 0.75 to 1.9 Rayleighs/$10^{20}$ cm$^{-2}$ for the
lines of sight studied by Reynolds, so the uncertainty in the conversion factor
is large.  
The value adopted here is the same as that used by Lagache et al. (2000).
Figure 5 shows that the derived values of $\epsilon$(H$^+$) are consistent
with zero for all regions at all wavelengths.  We have checked the dependence
of derived 100 $\micron$ $\epsilon$(H$^+$) values on the H I cut, and find that
they do not change significantly when the cut is varied by $\pm$ 30\%.

Our derived values of $\epsilon$(H$^+$) at 100 $\micron$ are compared
with previous results for other regions of the sky in Table 7. 
All of the results listed are from analyses similar to that used here,
with either H$\alpha$ intensity or centimeter wavelength radio continuum
intensity used as the tracer of the ionized gas.  Column 4 lists the electron density
adopted in each study for calculating the conversion from H$\alpha$ or radio continuum
intensity to $N$(H$^{+}$).
Uncertainty in this conversion may be as large as a factor of two or more for some
regions, but is not included in the uncertainty listed for $\epsilon$(H$^+$).
The emissivity values are shown plotted as a function of electron density in
Figure 8.  The 2$\sigma$ upper limits for our regions are comparable to that
obtained by Arendt et al. (1998) for the $10\arcdeg \times 12\arcdeg$ region 
at $l=144\arcdeg, b=-21\arcdeg$.
For the other previously studied regions, 100 $\micron$ emission was detected from 
the ionized gas component, and the derived $\epsilon$(H$^+$) is greater 
than or equal to the 100 $\micron$ $\epsilon$(H$^0$).  This can be
explained if the dust-to-gas mass ratio is the same in the ionized and neutral atomic 
components, with cases of enhanced emissivity in the ionized component
caused by Lyman alpha heating (for the Barnard's Loop region, Heiles et al. 2000) or
a local source of heating (for the Spica region, Boulanger et al. 1995, 
Zagury, Jones, and Boulanger 1998), or both (for the Galactic plane regions, 
Sodroski et al. 1997).  With the exception of the Q2 region,
the ionized gas in these regions is not representative of the general warm ionized 
medium observed at high latitudes. The electron density is an order of magnitude greater, 
and the ionized gas is much closer to the Galactic midplane ($z <$ 100 pc, 
compared to an exponential $z$ scale height of about 900 pc for the warm ionized medium
(Reynolds 1993)). 

The 100 $\micron$ $\epsilon$(H$^+$) value determined by Lagache et al. (2000) for the 
Q2 region without an H~I cut is not consistent with our result for the Q2 region 
with an H~I cut applied, or
with results for the other high latitude regions that sample the general warm
ionized medium.  We have applied our three-component decomposition to the full Q2 region 
(with no H~I cut) at 100 $\micron$ at the 1 degree resolution of the WHAM data.
This gave $\epsilon$(H$^+$) = 8.5 $\pm$ 1.0 nW~m$^{-2}$~sr$^{-1}$/$10^{20}$~cm$^{-2}$,
which is comparable to the upper limits in figure 8, and $\epsilon$(H$^0$) = 18.7~$\pm$~0.4 
nW m$^{-2}$ sr$^{-1}$/$10^{20}$ cm$^{-2}$.
(If an extinction correction is applied to the H$\alpha$ data as described in \S3.1, 
the derived $\epsilon$(H$^+$) is 5.5 $\pm$ 1.0 nW m$^{-2}$ sr$^{-1}$/$10^{20}$ cm$^{-2}$.)
The difference between these results and those of Lagache et al. is probably due to
the different angular resolution used. Their analysis used data for 122 positions at 7 degree 
FIRAS resolution, but the data were
oversampled and there are only about 30 independent FIRAS pointings within the region.
Our analysis used data for 1762 independent WHAM pointings that are within 3.5 degrees
(half of the FIRAS beam width) of any of their 122 positions. It is possible that some 
unmodeled
effect, such as dust associated with molecular gas, optical depth in the 21-cm line, or
variation of $\epsilon$(H$^0$), happens to correlate with 
the H$\alpha$ data at FIRAS resolution, but does not correlate as well at WHAM resolution.
The $\nu I_{\nu}(100 \micron)$ -- {\it N}(H~I) relation for the region shows curvature, which 
suggests that molecular gas or 21-cm line opacity may be present.  We find that 40\% of the 
region has temperature-corrected 100 $\micron$ intensity greater than the 3 MJy sr$^{-1}$ threshold
for significant fractional H$_2$ abundance (Gillmon and Shull 2006).
For these reasons, we consider the emissivity per H$^{+}$ ion derived for the full Q2 region,
from either our analysis or Lagache et al.'s analysis, to be of questionable reliability.

\section{Discussion}

  Our derived emissivity values for the ionized medium are statistically consistent 
with zero.  From Table 7, our derived 2$\sigma$ upper limits on 100 $\micron$ emissivity 
per H$^+$ ion are 0.33, 1.11, 0.37, 0.57, and 0.30 of the 100 $\micron$ emissivity per 
H atom for the HQBN, LH, NEP, Q1, and Q2 regions, respectively.  We adopt 0.4 as a 
representative upper limit on $\epsilon($H$^+)/\epsilon($H$^0)$ at 100 $\micron$ for these 
regions.  Possible explanations for this low value
include a lower dust-to-gas mass ratio or a weaker radiation field in the ionized medium
than in the neutral medium,
a difference in grain optical properties or grain size distribution, an error in our
adopted $I$(H$\alpha$)/$N$(H$^{+})$ conversion factor, or a shortcoming of using H$\alpha$
as a tracer of infrared emission from dust in the ionized medium.  In this section,
we discuss some of these possibilities.  Based on available observations and models,
it appears that our low derived $\epsilon($H$^+)/\epsilon($H$^0)$ ratio is partly due to a
lower dust-to-gas mass ratio in the WIM and partly due to error in using H$\alpha$ as a tracer, 
but one or more of the other factors may also contribute.
We discuss possible implications for derived CIB results in \S5.4.

\subsection{Dust-to-gas Mass Ratio}
A lower dust-to-gas mass ratio would be expected if the ionized gas in these regions has 
greater $z$ extent and lower density than the neutral gas, as is typical for the WIM in 
the solar vicinity (Reynolds 1991a). 
From interstellar absorption line observations, it has been inferred that abundances of 
heavy elements in the form of dust decrease with increasing $z$, and also decrease with 
decreasing gas density (e.g., Savage and Sembach 1996).  Most of these
studies have pertained to the neutral atomic medium, but from observations
of \break Al III and S III absorption lines, Howk and Savage (1999) found evidence that 
about 60-70\% of aluminum atoms in the WIM are in dust, compared to about 
90\% of Al in dust in H~II regions at low $z$.
This result is based on observations of two lines of sight that sample the 
WIM up to $z$ distances of 690 pc and 2800 pc, and four lines of sight through low density 
($n_e \sim$ 0.2 to 4 cm$^{-3}$) H~II regions at $z <$ 200 pc.  Howk and Savage also noted 
that the dust phase
Al abundance they obtained for the low $z$ H~II regions is comparable to previous, somewhat 
uncertain determinations for the warm neutral medium at low $z$.  Assuming that this low $z$
dust phase Al abundance is valid for the neutral atomic medium in the regions studied here, 
and that 100 $\micron$ emissivity per H nucleus varies in proportion to Al dust phase abundance,
one would expect $\epsilon$(H$^+$) to be about 20--30\% lower than $\epsilon$(H$^0$) at
100 $\micron$.\break
Thus the Howk and Savage results suggest that the difference in dust abundance between the
neutral and ionized components is not great enough to fully explain our derived 
upper limit on $\epsilon($H$^+)/\epsilon($H$^0)$.
To confirm this, dust phase Al abundance determinations 
for additional lines of sight through the WIM would be of interest, as would calculations of 
the relation between Al dust phase abundance and 100 $\micron$ emissivity for dust grain models.

\subsection{Interstellar Radiation Field}
  The lower 100 $\micron$ emissivity derived for the ionized medium could also be explained if 
it were subject to a weaker interstellar radiation field.  Heating of dust by Lyman alpha is 
expected to be negligible at the 
electron density estimated for the WIM (Spitzer 1978, Heiles et al. 2000). Also, there are no 
early type stars that are close enough to the lines of sight through our regions to cause
significant dust heating relative to that of the general interstellar radiation field.
Models of the spatial distribution of the interstellar radiation field at visual 
and ultraviolet wavelengths predict that the mean intensity increases with increasing $z$ up to 
about 200 pc, as light from distant stars in the Galactic disk becomes less attenuated, and 
then decreases with further increase in $z$ due to geometrical dilution.  
Wakker and Boulanger (1986) used their model of the radiation field to calculate the expected 
100 $\micron$ intensity of a diffuse cloud with a mixture of silicate and graphite grains.
The intensity was calculated for different distances of the cloud along two high latitude 
lines of sight (toward $b=90\arcdeg$, and toward $l=180\arcdeg$, $b=60\arcdeg$).  They found 
that the 100 $\micron$ \break
intensity varies by less than $\pm$20\% over the range in cloud height from $z =$ 0 -- 1 kpc.
Thus, it appears that the radiation field does not vary enough to explain our derived limit on
$\epsilon($H$^+)/\epsilon($H$^0)$, even 
if all the H~I gas were located near the maximum of the radiation field and all of the 
H~II gas were at $z =$ 1 kpc.

\subsection{$I$(H$\alpha$)/$N($H$^{+})$ Conversion Factor}
To change our derived upper limit on $\epsilon($H$^+)/\epsilon($H$^0)$ at 100 $\micron$ from 
0.4 to 1.0, the adopted $I$(H$\alpha$)/$N($H$^{+})$ conversion factor would need to be changed 
from 1.15 to 2.9 Rayleighs/$10^{20}$ cm$^{-2}$.
This is significantly larger than the largest value of \break 1.9 Rayleighs/$10^{20}$ cm$^{-2}$
measured by Reynolds (1991b) for lines of sight to four pulsars at high $z$.  Values 
obtained by Arendt et al. (1998) for an additional 5 pulsar lines of sight range from
0.8 to 1.2 Rayleighs/$10^{20}$ cm$^{-2}$.  Thus, it appears that there is not an error 
in the adopted conversion factor value that is large enough to fully explain the low
emissivity derived for the ionized medium. 

Independent evidence that supports our adopted conversion factor comes from interstellar 
absorption line studies.  Our adopted value corresponds to an effective electron density of 
0.08 cm$^{-3}$.  Electron density 
estimates from observations of absorption lines of excited C$^+$ toward extragalactic objects 
and high $z$ stars are comparable to this.  In the most extensive study to date, Lehner, 
Wakker, and Savage (2004) 
presented results 
for 43 such lines of sight at $\vert \thinspace b \thinspace \vert > 30 \arcdeg$. 
Most of the observed absorption line components are at low velocity.  For these, they find 
a mean density of $\langle n_e \rangle = 0.08 \pm 0.04$ cm$^{-3}$ (1$\sigma$ dispersion).  
For the Intermediate 
Velocity Arch, they find $\langle n_e \rangle = 0.03 \pm 0.01$ cm$^{-3}$, probably lower 
because the gas is at higher $z$ ($\sim1$ kpc).  The derived n$_e$ values are averages over 
C$^+$ regions in both the warm ionized medium and the warm neutral medium, but Lehner et al. 
used WHAM H$\alpha$ data to estimate that at least 
50\% of the excited C$^+$ column density originates in the WIM for an average line of sight.

\subsection{H$\alpha$ as a Tracer of Infrared Emission}
Our derived emissivity values and CIB results are subject to error if the WHAM H$\alpha$ 
data are not a good tracer of far infrared emission from the WIM in the regions we study.  
Far infrared intensity is proportional to dust column
density, whereas H$\alpha$ intensity is proportional to the square of
the ionized gas density integrated along the line of sight, $I$(H$\alpha$)
$\propto T_e^{-0.92} \int n_e^2 ds$, where $n_e$ is electron density
and $T_e$ is electron temperature (Reynolds 1992).
Thus errors would be expected in our results if the spatial variation of H$\alpha$ 
intensity for a region is caused more by differences in mean electron
density or mean electron temperature for different lines of sight than by 
differences in ionized gas column density.  
The approximate csc $\vert b \vert$ dependence of high latitude WHAM data (Haffner 
et al. 2003) provides evidence that H$\alpha$ is a reasonable tracer of $N$(H$^+$) 
on large angular scales, but this isn't necessarily true on the scales within the 
regions studied here.

Haffner, Reynolds, and Tufte (1999) and Reynolds, Haffner, and Tufte (1999) have
found evidence that variations in H$\alpha$ intensity may be largely due to variations
in electron density, based on their observations of H$\alpha$, [N II] $\lambda$ 6583, and 
[S II] $\lambda$ 6716 line intensities in the region $123\arcdeg < l < 164\arcdeg$, 
$-35\arcdeg < b < -6\arcdeg$, and 
previous observations of these lines in halos of edge-on galaxies.
The [N II]/H$\alpha$ and [S II]/H$\alpha$ intensity ratios are observed to increase
with increasing $z$, while the [S II]/[N II] ratio is nearly constant.
Interpreting the variations in [N II]/H$\alpha$ as primarily due to variations in electron
temperature, Haffner et al. and Reynolds et al. inferred $T_e$ values ranging from 6000 K 
to 11000 K, with temperature increasing as H$\alpha$ intensity decreases.  
They showed that this anticorrelation can be explained if variations in H$\alpha$ are 
largely due to variations in mean electron density, and a supplemental source of 
gas heating is present that dominates over photoionization at low density, causing
temperature to increase with decreasing density.  A number of possible supplemental
heating mechanisms have been proposed, and Reynolds et al. estimate the heating rate that
would be needed to explain the observations for each mechanism.

If H$\alpha$ is not a good tracer of ionized gas column density, our method of analysis tends 
to underestimate the infrared emissivity per H$^{+}$ ion and overestimate the CIB.  To show 
this, we consider a simple model in which the H$\alpha$ variation in a region is partly due 
to variation of $N$(H$^{+}$) and partly due to variation of effective electron density $n_{eff}$,
\begin{equation}
\frac {I(\rm H \alpha)} {\langle I(\rm H \alpha) \rangle} = 
\frac {N(\rm H^{+})} {\langle N(\rm H^{+}) \rangle} \frac {n_{eff}} {\langle n_{eff} \rangle}
\end{equation}
where $\langle I(H\alpha) \rangle$, $\langle N($H$^{+}) \rangle$, and $\langle n_{eff} \rangle$
are averages over all lines of sight through the region.  This assumes that electron 
temperature is constant, and $\langle N($H$^{+})\enspace n_{eff} \rangle = 
\langle N($H$^{+}) \rangle \thinspace \langle n_{eff} \rangle$.
We adopt
\begin{equation}
\frac {N(\rm H^{+})} {\langle N(\rm H^{+}) \rangle} = 
\left[\frac {I(\rm H\alpha)} {\langle I(\rm H\alpha) \rangle} \right]^{p} ,
\end{equation}
and
\begin{equation}
\frac {n_{eff}} {\langle n_{eff} \rangle} = \left[\frac {I(\rm H\alpha)} 
{\langle I(\rm H\alpha) \rangle} \right]^{1-p} ,
\end{equation}
where $p$ is the fraction of the variation of log $I$(H$\alpha$) that is caused by variation of $N$(H$^+$).
We assume that, averaged over large angular scales, $N$(H$^+$) and $I$(H$\alpha$) are directly proportional,
\begin{equation}
\langle N(\rm H^{+}) \rangle = \langle I(\rm H\alpha) \rangle /\it c ,
\end{equation}
where $c$ is our adopted conversion factor of 1.15 Rayleighs/$10^{20}$ cm$^{-2}$.
The infrared emissivity per H nucleus is assumed to be constant within each gas 
phase, so the infrared emission from each phase is proportional to its gas column density.
The infrared emission from the ionized phase is
\begin{equation}
\nu I_{\nu}^{WIM} = \epsilon (\rm H^{+}) \it N(\rm H^{+}) 
 = \epsilon (\rm H^{+}) \langle \it I(\rm H\alpha) \rangle^{1-{\it p}} \it I(\rm H\alpha)^{\it p} /\it c.
\end{equation}
Assuming the distributions of $N$(H~I) and $I$(H$\alpha$) are uncorrelated, the H$\alpha$
coefficient $B_1$ that would be obtained from our decomposition is the mean slope of
the $\nu I_{\nu}^{WIM}$ -- $I$(H$\alpha$) relation.  If the distribution of H$\alpha$
intensities is symmetric about the mean, this is given by
\begin{equation}
B_1 = \left. \frac {d \thinspace \nu I_{\nu}^{WIM}} {d \thinspace I(\rm H\alpha)} \right|_{I(\rm H\alpha)
\thinspace = \thinspace \langle I(\rm H\alpha) \rangle} = p \epsilon(\rm H^{+}) /\it c
\end{equation}
and the emissivity per H$^+$ ion derived from the analysis underestimates the actual
emissivity,
\begin{equation}
\epsilon_{derived} (\rm H^{+}) = \it c B_1 = p \epsilon (\rm H^{+}) .
\end{equation}
Thus, our derived limit $\epsilon$(H$^+$)/$\epsilon$(H$^0) < 0.4$ may be consistent with no 
real difference in emissivity between the ionized and neutral phases if only a small
fraction of the H$\alpha$ variation in our regions is caused by
variation of $N$(H$^{+}$), i.e., if $p < 0.4$ in the context of this simple model.
Berkhuijsen, Mitra, and M\"uller (2006) have estimated $p = 0.68 \pm 0.04$
for the high latitude diffuse ionized gas, using pulsar dispersion measure data and 
WHAM H$\alpha$ data toward a sample of 157 pulsars at $\vert b \vert > 5\arcdeg$ 
(see their Figure 7b).  
This result suggests that our low derived $\epsilon$(H$^+$)/$\epsilon$(H$^0$) is at least partly
due to error in using H$\alpha$ as a tracer.  

  The overestimate of the CIB for our simple model is given by the intercept of a linear fit 
to the $\nu I_{\nu}^{WIM}$ -- $I$(H$\alpha$) relation.  To a good approximation, this intercept
is given by 
\begin{equation}
\Delta \thinspace (\nu I_{\nu}) = \left. \nu I_{\nu}^{WIM} \right|_{I(\rm H\alpha) \thinspace = \thinspace
\langle I(\rm H\alpha) \rangle} - B_1 \langle I(\rm H\alpha) \rangle .
\end{equation}
Using equations (6), (7), and (8), we obtain
\begin{equation}
\Delta \thinspace (\nu I_{\nu}) = (1-p) \thinspace \epsilon (\rm H^{+}) \thinspace \langle \it N(\rm H^{+}) \rangle .
\end{equation}
We have used this result to estimate the possible effect of emission from the WIM that is not 
correlated with H$\alpha$ or $N$(H~I)
on the CIB results derived from our three-component fits. 
We subtracted $\Delta \thinspace (\nu I_{\nu})$ offsets from the derived residual intensities $C_1$ 
for each region
assuming $p = 0.5 \pm 0.5$, $\langle N($H$^{+}) \rangle$ = $\langle I($H$\alpha) \rangle$/(1.15
Rayleighs/$10^{20}$ cm$^{-2}$), and
$\epsilon ($H$^{+}) = (0.5 \pm 0.5) \thinspace \epsilon ($H$^{0})$, using values of $A_1$ in Table 3
for $\epsilon ($H$^{0})$.  For each wavelength, 
we then calculated the weighted average of the reduced residual intensity over the five regions, 
treating the $\pm 0.5$ uncertainties in $p$ and $\epsilon ($H$^{+}) / \epsilon ($H$^{0})$ as 
additional independent sources of error.  The resulting average residual intensities are
$21 \pm 27$, $17.0 \pm 6.3$, $23.5 \pm 8.5$, and $12.0 \pm 2.9$ nW m$^{-2}$ sr$^{-1}$ at 60,
100, 140, and 240 $\micron$, respectively.  At 140 and 240 $\micron$, these results are only 6\% lower
than the results from our three-component fits.

  Equation (11) is the same as the expression for the CIB overestimate from a two-component fit 
(using H~I without H$\alpha$) where $p$ is the fraction of $N$(H$^{+}$) that is correlated with $N$(H~I). 
Thus we can use the same procedure to estimate the possible effect of emission from the WIM that is not
correlated with $N$(H~I) on the CIB results of Hauser et al. (1998).  We make the same assumptions as in 
the previous paragraph for $p$, $\langle N($H$^{+}) \rangle$, and $\epsilon ($H$^{+})$.  The HQBS region 
included in the Hauser et al. 
analysis is below the declination limit of the WHAM survey, so for this region we used data from the 
Southern H-Alpha Sky Survey of Gaustad et al. (2001) as processed by Finkbeiner (2003).
We find that the weighted-average residual intensities from the Hauser et al.
analysis would be reduced to $20.9 \pm 6.2$, $23.6 \pm 7.0$, and $12.9 \pm 2.5$ nW m$^{-2}$ sr$^{-1}$ 
at 100, 140, and 240 $\micron$, respectively.  These results are only about 5\% lower than those of
Hauser et al.  

  The possible effect of the ionized medium is also estimated to be small for the 125-2000~$\micron$ 
CIB spectrum determined by Fixsen et al. (1998) from FIRAS observations. In one of their analyses, 
Galactic emission template maps were constructed from DIRBE 140~$\micron$ and 240~$\micron$ data with 
the Hauser et al. (1998) CIB and zodiacal light subtracted, and the CIB spectrum was obtained 
by correlating FIRAS data with these templates.   Our estimate of $\sim5$\% for the possible error 
of the DIRBE CIB results also applies to the $140 \thinspace \lsim \thinspace \lambda \thinspace \lsim
\thinspace  240 \thinspace \micron$ part of the 
CIB spectrum derived from this analysis.  At longer wavelengths, the error is expected to decrease. 
This is because the spectrum of emission from the ionized medium is expected to be similar to the 
spectrum of emission from the H I phase, and
the ratio of this spectrum to the CIB spectrum decreases with increasing wavelength for 
$\lambda > 240 \thinspace \micron$.   Since all three of the Fixsen et al.
analyses gave consistent results for the CIB spectrum, the possible error due to the WIM
should be small for their average CIB spectrum from the three methods.

\section{Summary}

  We used WHAM H$\alpha$ data as a tracer of far infrared emission from 
the warm ionized phase of the ISM in an effort to determine the intensity
of this emission at high Galactic latitudes and to assess its effect on
determination of the cosmic far infrared background.  We studied five
high latitude regions, including regions previously
analyzed by Hauser et al. (1998) and Lagache et al. (2000).  For each region, 
we decomposed {\it COBE}/DIRBE data at 60, 100, 140, and 240 $\micron$ into a 
sum of an H~I correlated component, an H$\alpha$ correlated component, 
and a residual component. Uncertainties in our results due to omission of an 
H$_2$ correlated component are expected to be negligible, based on results of
a {\it FUSE} high latitude H$_2$ absorption line survey.
We found that the intensity of the H$\alpha$ correlated component is consistent
with zero within the uncertainties for all regions at all wavelengths.
From the mean intensities of the residual components, we derived estimates of the
CIB at 140 and 240 $\micron$ and upper limits to the CIB at 60 and 100 $\micron$ (Table 4).
We repeated the analysis without including an H$\alpha$ correlated component, and 
the derived CIB results did not change significantly.
Our CIB estimates and upper limits are similar to previous CIB determinations 
for which the FIR emission from the ISM was traced only by H~I column density.
We conclude that addition of an H$\alpha$ correlated component in modeling the
ISM emission at high Galactic latitude has negligible effect on derived CIB results.
We did not perform detailed anisotropy tests on the maps of residual intensity
from our analysis, so our CIB results do not supersede the results of Hauser et al. 
(1998).  

We derived 2$\sigma$ upper limits to the 100 $\micron$ emissivity per H$^+$ ion
for the five regions that are typically about 40\% of the emissivity per H atom 
for the neutral atomic medium.  Available evidence suggests that this low value is
partly due to a lower dust-to-gas mass ratio in the ionized medium than in the
neutral atomic medium, and partly due to a shortcoming of using H$\alpha$ as a 
tracer of FIR emission, which causes our analysis to underestimate the emissivity
of the ionized medium.  Other possible effects that may play a role include
a weaker radiation field in the ionized medium than in the neutral medium,
a difference in grain optical properties or grain size distribution, or
an error in our adopted $I$(H$\alpha$)/$N$(H$^{+}$) conversion factor.
(The value of 100 $\micron$ emissivity per H$^+$ ion derived by Lagache et al. (2000)
is much greater than the upper limit we derived for their region.   Our analysis
differs from theirs in that (1) we exclude positions where $N$(H~I) is greater than 
$3 \times 10^{20}$ cm$^{-2}$, where emission from H$_2$ associated 
dust or 21-cm line opacity may be significant, and (2) we analyze 
data at a resolution of 1$\arcdeg$ instead of 7$\arcdeg$.)

H$\alpha$ observations have previously been used in this kind of analysis with
apparent success for ionized regions that have higher density and are at low $z$.
However, for the general high latitude WIM, evidence from Reynolds, Haffner, and Tufte 
(1999) and Berkhuijsen, Mitra, and M\"uller (2006) suggests that variations in H$\alpha$ 
are not entirely due to variations in H$^+$ column density, but are also due to 
differences in mean electron density for different lines of sight.  Thus, H$\alpha$
may not be an accurate tracer of far infrared emission from the WIM.

If H$\alpha$ intensity is not a good tracer, any emission from the WIM
that is not correlated with H$\alpha$ or H~I would contribute to our
derived residual component for each region, so our analysis could overestimate the CIB.
The agreement with the Hauser et al. (1998) results could be because their analysis 
overestimates the CIB by a similar amount. 
We used WHAM data to estimate the possible effect on our CIB results
and on the Hauser et al. results, assuming that the mean H$\alpha$ intensity for
each region can be used to estimate its mean H$^+$ column density.  In each case, 
we estimated the effect to be 
only about 5\% at \break 140 and 240 $\micron$, which is much smaller than the quoted 
uncertainties.  The possible effect of emission from the WIM is also estimated to
be small for the 125-2000 $\micron$ CIB spectrum determined by Fixsen et al. (1998)
from FIRAS observations.

We estimated upper limits on a possible diffuse component of the CIB by comparing 
the Hauser et al. (1998) CIB results with the integrated galaxy brightness determined 
by Dole et al. (2006) from {\it Spitzer} source counts.  We obtained 2$\sigma$ upper 
limits on a diffuse component of 26 nW m$^{-2}$ sr$^{-1}$ at 140 $\micron$ and 
8.5~nW~m$^{-2}$~sr$^{-1}$ at 240 $\micron$.

\bigskip

  We thank F. Boulanger, J. C. Howk, and G. Lagache for useful discussions, and G. Lagache for
providing us with data from the previous study of the Q2 region.  We thank the referee for
helpful comments.  This paper is dedicated to 
the memory of our friend and colleague Thomas J. Sodroski, and we acknowledge his 
contributions in the formative stages of this work.  This research was supported by 
NASA Astrophysical Data Program NRA 99-01-ADP-137.

\clearpage

\begin{deluxetable}{cccc}
\tabletypesize{\scriptsize}
\tablewidth{0pt}
\tablecaption{Datasets}
\tablehead{
\colhead{} &
\colhead{} &
\colhead{Velocity} &
\colhead{}\\
\colhead{Dataset} &
\colhead{Resolution} &
\colhead{Integration} &
\colhead{Reference}
}
\startdata 
DIRBE Zodi--Subtracted    & $0\fdg7$    &  \nodata  &    Hauser et al. (1997)\\
Mission Average Maps & & & \\
 & & & \\
Leiden--Dwingeloo  &   $0\fdg6$  &  $-450 < v_{\rm LSR} < 400$ km s$^{-1}$  &  Hartmann \& Burton (1997)\\
H I Survey            & & & \\     
 & & & \\
Lockman Hole     &        $21'$  &     $-150 <  v_{\rm LSR} < 100$ km s$^{-1}$  &  Snowden et al. (1994)\\
H I map & & & \\
 & & & \\
NEP H I Map        &      $21'$   &    $-150 <  v_{\rm LSR} < 150$ km s$^{-1}$   & Elvis et al. (1994)\\
 & & & \\
WHAM H$\alpha$     &        $1\fdg0$   &  $-80 <  v_{\rm LSR} <  80$ km s$^{-1}$   & Haffner et al. (2003)\\
Sky Survey & & & \\
\enddata
\end{deluxetable}

\begin{deluxetable}{ccccc}
\tabletypesize{\scriptsize}
\tablewidth{0pt}
\tablecaption{Test of Analysis for Q2 Region}
\tablehead{
\colhead{Wavelength} &
\colhead{$A_1$\tablenotemark{a}} &
\colhead{$B_1$} &
\colhead{$C_1$} &
\colhead{Source}\\
\colhead{($\micron$)} &
\colhead{(nW m$^{-2}$ sr$^{-1}$/10$^{20}$ cm$^{-2}$)} &
\colhead{(nW m$^{-2}$ sr$^{-1}$/ Rayleigh)\tablenotemark{b}} &
\colhead{(nW m$^{-2}$ sr$^{-1}$)} &
\colhead{}
}
\startdata 
     100    &     $14.9\pm0.1$  &   $14.0\pm0.4$  &  $23.4\pm6.3$  &  Lagache et al. (2000)\\
     100    &     $14.9\pm1.1$  &  $14.0\pm3.0$  &  $23.4\pm3.8$   & This paper\\
& & & & \\

     140    &     $20.0\pm0.6$  &   $23.4\pm1.9$  &  $24.3\pm11.6$  &  Lagache et al. (2000) \\
     140    &     $20.0\pm2.0$  &   $23.4\pm5.9$  &  $24.3\pm7.9$ &   This paper\\
& & & & \\

     240    &     $ 9.6\pm0.3$ &   $12.7\pm0.8$  &  $11.0\pm6.9$  &  Lagache et al. (2000)\\
     240    &      $9.6\pm1.3$ &    $12.7\pm3.8$  &  $11.0\pm5.1$  &  This paper\\
\enddata
\tablenotetext{a}{Parameter values from fits of the form  $\nu I_{\nu}(\lambda) = A_1 N({\rm H I}) + B_1 I({\rm H}\alpha) + C_1$}
\tablenotetext{b}{1 Rayleigh = 10$^{6}$/4$\pi$ photons s$^{-1}$ cm$^{-2}$ sr$^{-1}$ = 0.24 nW m$^{-2}$ sr$^{-1}$ at H$\alpha$}
\end{deluxetable}

\begin{deluxetable}{cccccccc}
\tabletypesize{\tiny}
\tablewidth{0pt}
\tablecaption{Decomposition Results}
\tablehead{
\colhead{Region}&
\colhead{DIRBE Band} &
\colhead{Fit type}&
\colhead{$A_i$} &
\colhead{$B_i$} &
\colhead{$C_i$} &
\colhead{$N_{\rm points}$}&
\colhead{$\chi^2_{\nu}$}\\
\colhead{}&
\colhead{($\micron$)} &
\colhead{$i$}&
\colhead{(nW m$^{-2}$ sr$^{-1}$/10$^{20}$ cm$^{-2}$)} &
\colhead{(nW m$^{-2}$ sr$^{-1}$/ Rayleigh)\tablenotemark{c}} &
\colhead{(nW m$^{-2}$ sr$^{-1}$)} &
\colhead{}&
\colhead{}
}
\startdata
 HQBN &  60  & 1\tablenotemark{a} &  $  8.4\pm0.6$ &  $  0.5\pm2.3$ &  $ 18.0\pm1.0$  &   535  &   1.25\\
      &      & 2\tablenotemark{b} &  $  8.4\pm0.5$ &    \nodata     &  $ 18.2\pm0.6$  &   535  &   1.25\\
      & 100  & 1 &  $ 23.1\pm1.2$ &  $ -5.3\pm5.9$ &  $ 16.8\pm2.3$  &   535  &   2.11\\
      &      & 2 &  $ 22.7\pm0.9$ &    \nodata     &  $ 14.8\pm1.3$  &   535  &   2.12\\
      & 140  & 1 &  $ 19.6\pm7.1$ &  $-30.8\pm30.0$ &  $ 36.0\pm12.9$  &   535  &   1.31\\
      &      & 2 &  $ 17.3\pm5.5$ &    \nodata     &  $ 23.8\pm7.7$  &   535  &   1.32\\
      & 240  & 1 &  $  8.4\pm2.1$ &  $  1.8\pm8.5$ &  $ 11.2\pm3.6$  &   535  &   1.12\\
      &      & 2 &  $  8.6\pm1.7$ &    \nodata     &  $ 11.9\pm2.3$  &   535  &   1.12\\
& & & & & & & \\
 LH   &  60  & 1 &  $  4.9\pm1.1$ &  $  4.8\pm2.4$ &  $ 24.2\pm1.0$  &   215  &   0.86\\
      &      & 2 &  $  5.3\pm0.9$ &    \nodata     &  $ 25.0\pm0.9$  &   215  &   0.91\\
      & 100  & 1 &  $ 18.8\pm2.0$ &  $  7.7\pm5.1$ &  $ 19.3\pm2.0$  &   215  &   2.20\\
      &      & 2 &  $ 20.2\pm1.7$ &    \nodata     &  $ 19.9\pm1.7$  &   215  &   2.27\\
      & 140  & 1 &  $ 18.6\pm11.8$ & $ -8.6\pm39.6$ &  $ 24.6\pm13.1$  &   246  &   1.34\\
      &      & 2 &  $ 17.7\pm9.4$ &    \nodata     &  $ 23.2\pm10.6$  &   246  &   1.34\\
      & 240  & 1 &  $  8.9\pm3.4$ &  $-18.0\pm10.6$ &  $ 17.7\pm3.8$  &   246  &   1.02\\
      &      & 2 &  $  7.3\pm2.8$ &    \nodata     &  $ 14.7\pm3.2$  &   246  &   1.05\\
& & & & & & & \\
 NEP  &  60  & 1 &  $  8.7\pm0.8$ &  $ -0.1\pm1.2$ &  $ 14.1\pm2.8$  &    45  &   1.12\\
      &      & 2 &  $  8.7\pm0.6$ &    \nodata     &  $ 14.2\pm2.2$  &    45  &   1.10\\
      & 100  & 1 &  $ 18.9\pm2.0$ &  $  0.7\pm2.7$ &  $ 16.4\pm6.6$  &    45  &   2.86\\
      &      & 2 &  $ 19.3\pm1.3$ &    \nodata     &  $ 16.0\pm5.2$  &    45  &   2.81\\
      & 140  & 1 &  $  8.9\pm8.2$ &  $  9.0\pm11.5$ &  $ 54.8\pm29.8$  &    54  &   1.02\\
      &      & 2 &  $ 13.6\pm5.5$ &    \nodata     &  $ 50.5\pm23.9$  &    54  &   1.04\\
      & 240  & 1 &  $  4.6\pm2.8$ &  $  5.8\pm4.2$ &  $ 19.9\pm10.3$  &    54  &   1.10\\
      &      & 2 &  $  7.5\pm2.0$ &    \nodata     &  $ 17.5\pm8.6$  &    54  &   1.22\\
& & & & & & & \\
 Q1   &  60  & 1 &  $  6.6\pm0.8$ &  $  1.7\pm1.4$ &  $ 17.8\pm2.0$  &   217  &   2.51\\
      &      & 2 &  $  6.9\pm0.6$ &    \nodata     &  $ 19.1\pm1.3$  &   217  &   2.53\\
      & 100  & 1 &  $ 20.8\pm1.7$ &  $  2.8\pm3.7$ &  $ 13.9\pm4.6$  &   225  &   4.22\\
      &      & 2 &  $ 21.1\pm1.3$ &    \nodata     &  $ 16.0\pm2.8$  &   225  &   4.23\\
      & 140  & 1 &  $ 24.1\pm7.0$ &  $-13.3\pm14.0$ &  $ 38.0\pm18.6$  &   234  &   1.33\\
      &      & 2 &  $ 22.4\pm5.5$ &    \nodata     &  $ 27.7\pm11.9$  &   234  &   1.34\\
      & 240  & 1 &  $ 11.0\pm2.3$ &  $  2.6\pm4.5$ &  $  8.5\pm6.3$  &   235  &   1.42\\
      &      & 2 &  $ 11.3\pm1.9$ &    \nodata     &  $ 10.5\pm4.0$  &   235  &   1.42\\
& & & & & & & \\
 Q2   &  60  & 1 &  $  5.4\pm0.5$ &  $  0.3\pm0.8$ &  $ 22.5\pm1.1$  &   499  &   2.61\\
      &      & 2 &  $  5.5\pm0.4$ &    \nodata     &  $ 22.6\pm0.8$  &   499  &   2.60\\
      & 100  & 1 &  $ 15.4\pm0.9$ &  $ -1.0\pm2.5$ &  $ 21.9\pm1.4$  &   487  &   6.63\\
      &      & 2 &  $ 15.2\pm0.6$ &    \nodata     &  $ 21.6\pm1.4$  &   487  &   6.63\\
      & 140  & 1 &  $ 24.5\pm4.8$ &  $ -0.9\pm9.4$ &  $ 15.3\pm8.6$  &   496  &   1.21\\
      &      & 2 &  $ 24.3\pm3.4$ &    \nodata     &  $ 14.9\pm7.4$  &   496  &   1.21\\
      & 240  & 1 &  $ 10.2\pm1.6$ &  $ -0.4\pm3.1$ &  $ 11.1\pm3.1$  &   497  &   1.33\\
      &      & 2 &  $ 10.1\pm1.2$ &    \nodata     &  $ 10.9\pm2.6$  &   497  &   1.33\\
\enddata
\tablenotetext{a}{Fit type 1:  $\nu I_{\nu}(\lambda) = A_1 N({\rm H I}) + B_1 I({\rm H}\alpha) + C_1$}
\tablenotetext{b}{Fit type 2:  $\nu I_{\nu}(\lambda) = A_2 N({\rm H I}) + C_2$}
\tablenotetext{c}{1 Rayleigh = 10$^{6}$/4$\pi$ photons s$^{-1}$ cm$^{-2}$ sr$^{-1}$ = 0.24 nW m$^{-2}$ sr$^{-1}$ at H$\alpha$}
\end{deluxetable}

\begin{deluxetable}{cccccc}
\tabletypesize{\scriptsize}
\tablewidth{0pt}
\tablecaption{CIB Limits and Detections\tablenotemark{a}}
\tablehead{
\colhead{ISM}&
\multicolumn{4}{c}{$\nu I_{\nu}({\rm CIB})$  (nW m$^{-2}$ sr$^{-1}$)}&
\colhead{}\\
\cline{2-5}
\colhead{Tracers}&
\colhead{60 $\micron$}&
\colhead{100 $\micron$}&
\colhead{140 $\micron$}&
\colhead{240 $\micron$}&
\colhead{Reference}
}
\startdata
H I                &  $< 75\ (20.6\pm27)$ & $< 34\ (21.9\pm6.1)$ & $   25.0\pm6.9  $ & $ 13.6\pm2.5  $ & Hauser et al. (1998)\\
H I                &  $< 75\ (20.5\pm27)$ & $< 30\ (18.2\pm6.1)$ & $ 22.0\pm7.0^b  $ & $12.1\pm2.5^b $ & This paper\\
H I \& H$\alpha$  &  $< 75\ (20.9\pm27)$ & $< 32\ (19.8\pm6.1)$ & $  25.1\pm8.0^b  $ & $12.8\pm2.8^b $ & This paper\\
H I \& H$\alpha$  &     \nodata       & $  23.4\pm6.3^b  $ & $< 47\ (24.2\pm11.6) $ & $< 25 (11.0\pm6.9) $ & Lagache et al. (2000)\\
\enddata
\tablenotetext{a}{Upper limits are given for cases where the residuals are not 3$\sigma$ 
greater than zero or where they are non-isotropic.}
\tablenotetext{b}{A test for anisotropy was not performed.}
\end{deluxetable}

\begin{deluxetable}{lcccc}
\tabletypesize{\scriptsize}
\tablewidth{0pt}
\tablecaption{Effect of Interplanetary Dust Model on CIB Determination}
\tablehead{
\colhead{} &
\multicolumn{4}{c}{$\nu I_{\nu}$  (nW m$^{-2}$ sr$^{-1}$)} \\
\cline{2-5}
\colhead{}&
\colhead{60 $\micron$}&
\colhead{100 $\micron$}&
\colhead{140 $\micron$}&
\colhead{240 $\micron$}
}
\startdata
Hauser et al. (1998) mean residual \tablenotemark{a}   & 20.6 $\pm$ 27 & 21.9 $\pm$ 6.1 & 25.0 $\pm$ 6.9 & 13.6 $\pm$ 2.5 \\
Wright (2004) mean residual \tablenotemark{b}        &   -8   $\pm$ 14 & 12.5 $\pm$ 5   & 22   $\pm$ 7   & 13   $\pm$ 2.5 \\
Residual difference (row 1 - row 2) &    28.6          &  9.4           &  3             &  0.6           \\
Zodi subtraction uncertainty \tablenotemark{c}        &    26.7          &  6.0           &  2.3           &  0.5           \\
\enddata
\tablenotetext{a}{Based on the IPD model of Kelsall et al. (1998).}
\tablenotetext{b}{Based on the IPD model of Gorjian et al. (2000).}
\tablenotetext{c}{1$\sigma$ uncertainty in zodiacal light subtraction for the Hauser et al. analysis, from Kelsall et al. (1998).}
\end{deluxetable}
 
\begin{deluxetable}{lcc}
\tabletypesize{\scriptsize}
\tablewidth{0pt}
\tablecaption{Comparison of Integrated Galaxy Brightness and CIB Measurements}
\tablehead{
\colhead{} &
\colhead{140 $\micron$} &
\colhead{240 $\micron$}
}
\startdata
Integrated galaxy brightness \tablenotemark{a}  \tablenotemark{b}      &  13.7 $\pm$ 1.7  & 10.7 $\pm$ 1.4 \\
DIRBE CIB \tablenotemark{a} \tablenotemark{c}                        &  25.0 $\pm$ 6.9  & 13.6 $\pm$ 2.5 \\
DIRBE CIB transformed to FIRAS scale \tablenotemark{a} \tablenotemark{c} &  15.0 $\pm$ 5.9  & 12.7 $\pm$ 1.6 \\
Integrated galaxy brightness/CIB (DIRBE scale)  &  0.55 $\pm$ 0.17  &  0.79 $\pm$ 0.18 \\
Integrated galaxy brightness/CIB (FIRAS scale)  &  0.91 $\pm$ 0.37  &  0.84 $\pm$ 0.15 \\
CIB (DIRBE scale) - Integrated galaxy brightness \tablenotemark{a} & $< 26\ (11.3 \pm 7.1)$  & $<8.5\ (2.9 \pm 2.8)$ \\
CIB (FIRAS scale) - Integrated galaxy brightness \tablenotemark{a} & $<14\ (1.3 \pm 6.1)$  & $< 6.2\ (2.0 \pm 2.1)$ \\
\enddata
\tablenotetext{a}{$\nu I_{\nu}$ in nW m$^{-2}$ sr$^{-1}$.}
\tablenotetext{b}{Dole et al. (2006) 160 $\micron$ result from Spitzer MIPS observations, scaled to DIRBE wavelengths.}
\tablenotetext{c}{From Hauser et al. (1998).}
\end{deluxetable}

\begin{deluxetable}{ccccccc}
\tabletypesize{\scriptsize}
\tablewidth{0pt}
\tablecaption{100 $\micron$ Emissivity Determinations}
\tablehead{
\colhead{Region} &
\colhead{l} &
\colhead{$\vert$b$\vert$} &
\colhead{Adopted} &
\colhead{100 $\micron$ Emissivity} &
\colhead{100 $\micron$ Emissivity} &
\colhead{Reference}\\
\colhead{} &
\colhead{} & 
\colhead{} &
\colhead{$n_e$ (cm$^{-3}$)} &
\colhead{per H$^{+}$ Ion\tablenotemark{a}} &
\colhead{per H atom\tablenotemark{a}} &
\colhead{}
}
\startdata
   HQBN   &    48-145  &  60-75  &  0.08 & $-6.1 \pm 6.8 $ & $ 23.1 \pm 1.2 $ &  This paper  \\
    LH    &   130-163  &  46-65  &  0.08 & $ 8.9 \pm 5.9 $ & $ 18.8 \pm 2.0 $ &  This paper  \\
   NEP    &    93-100  &  25-34  &  0.08 & $ 0.8 \pm 3.1 $ & $ 18.9 \pm 2.0 $ &  This paper  \\
    Q1    &    30- 80  &  30-41  &  0.08 & $ 3.2 \pm 4.3 $ & $ 20.8 \pm 1.7 $ &  This paper  \\
 Q2 (HI cut) & 97-170  &  28-52  &  0.08 & $-1.2 \pm 2.9 $ & $ 15.4 \pm 0.9 $ &  This paper  \\
 Q2 (no HI cut) &    97-170  &  22-52  &  0.08 & $16.2 \pm 0.5 $  & $ 14.9 \pm 0.1 $ &  Lagache et al. (2000) \\
{\it l} = 144, {\it b} = -21 & 136-150 &  16-27  &  0.09 & $-0.4 \pm 1.9 $ & $16.3 \pm  1.1 $  &  Arendt et al. (1998), this paper \\
  Spica   &   297-343  &  39-70  &  0.6  & $ 87 \pm 20 $  & $19.5 \pm 1.6 $ &  Arendt et al. (1998), this paper \\
Eridanus Superbubble & 180-210 & 15-50 & 0.8 & $ 20   $  &  $ 19          $  &  Heiles et al. (1999) \\
 Barnard's Loop & 208-217 & 13-27 & 2    & $ 34       $  &  $ 17          $  &  Heiles et al. (2000) \\
Outer Galactic Plane & 90-270 & 0-5 & 2  & $ 54 \pm  5 $  & $ 6.8 \pm 0.4 $  &  Sodroski et al. (1997) \\
Inner Galactic Plane & 270-90 & 0-5 & 10 & $380 \pm 17 $  & $ 37-180      $  &  Sodroski et al. (1997) \\
\enddata
\tablenotetext{a}{in nW m$^{-2}$ sr$^{-1}$/10$^{20}$ cm$^{-2}$}
\end{deluxetable}

\clearpage

\begin{figure}
\figurenum{1}
\epsscale{1.00}
\plotone{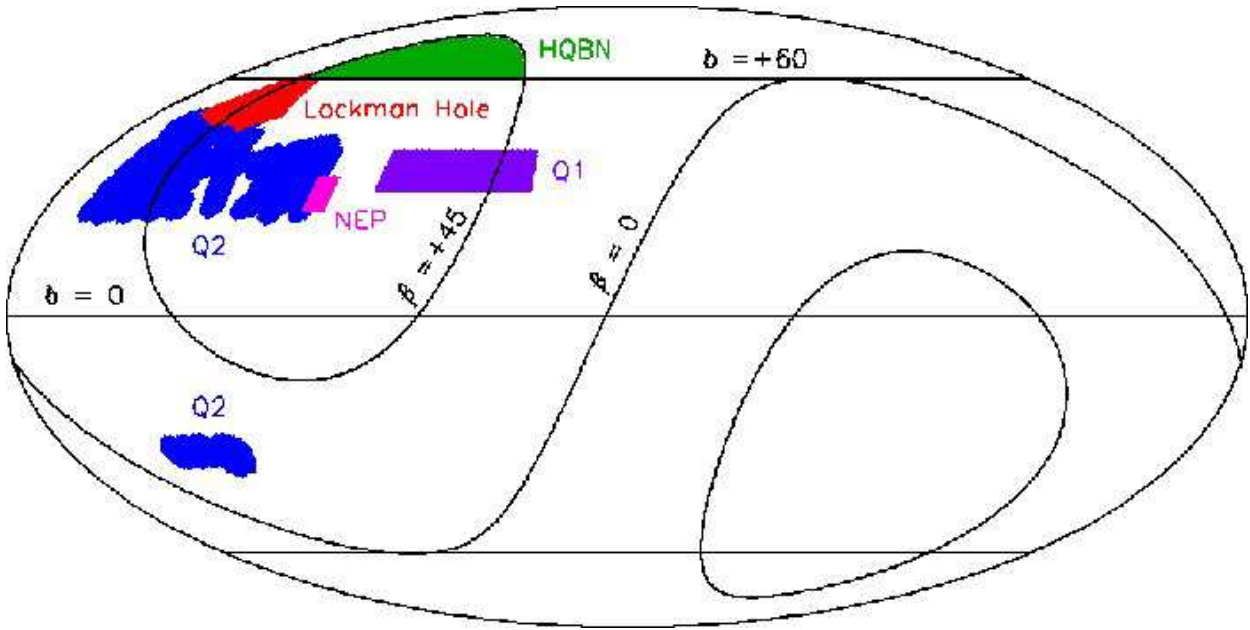}
\caption{Location of the regions analyzed on a Galactic coordinate
Mollweide projection centered at $l = 0\arcdeg$.  The Lockman Hole (LH) region 
is the region mapped in H I by Snowden et al. (1994).  The north ecliptic pole 
(NEP) region is the region mapped in H~I by Elvis, Lockman and Fassnacht (1994). 
The DIRBE high quality B north (HQBN) region is defined as the region at Galactic 
latitude $b > +60\arcdeg$ and ecliptic latitude $\beta > +45\arcdeg$.  
The first quadrant (Q1) region is defined by $30\arcdeg < l < 80\arcdeg, 
30\arcdeg < b < 41\arcdeg$.  The second quadrant (Q2) region is a region 
previously studied by Lagache et al. (2000).}
\end{figure}

\clearpage

\begin{figure}
\figurenum{2}
\epsscale{0.75}
\plotone{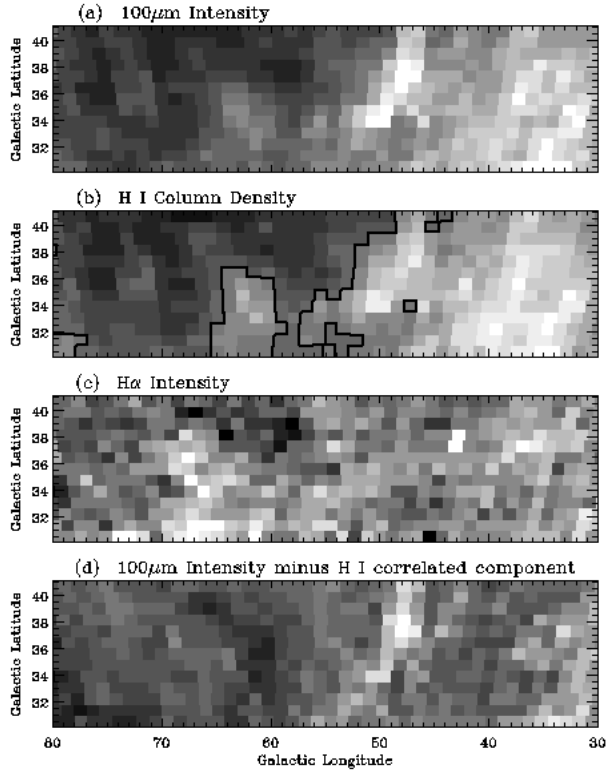}
\caption{Images of the Q1 region in (a) DIRBE 100 $\micron$ intensity 
after subtraction of the interplanetary dust emission model, (b) H~I column density
from the Leiden-Dwingeloo survey, (c) H$\alpha$ intensity from the WHAM
Northern Sky Survey, and (d) 100 $\micron$ intensity as in (a) but with 
the component that is correlated with H~I column density subtracted.
The distributions of H~I and H$\alpha$ differ from each other and differ
from an isotropic distribution, so the infrared data can be decomposed
into a sum of the three distributions.  No correlation is seen 
between H$\alpha$ and the residual 100 $\micron$ emission in (d),
consistent with the low value of the H$\alpha$ coefficient $B_1$ obtained from
our analysis.  Possibly the 100 $\micron$ emissivity of the ionized medium 
is low and the intensity variations in (d) are not related to the ionized
medium, or H$\alpha$ is not a good tracer of the 100 $\micron$ emission 
from this medium.  The image display ranges, from black to white, are 
20 to 170 nW m$^{-2}$ sr$^{-1}$ for (a), 0.4 to 6.3 10$^{20}$ atoms cm$^{-2}$ 
for (b), 0.5 to 1.6 R for (c), and -25 to 50 nW m$^{-2}$ sr$^{-1}$ for (d).
The black contour in (b) shows the {\it N}(H~I) cut of 3 10$^{20}$ atoms cm$^{-2}$ 
used in the analysis, as described in \S3.1.}
\end{figure}
\clearpage

\begin{figure} 
\figurenum{3}
\includegraphics[angle=90,width=\textwidth]{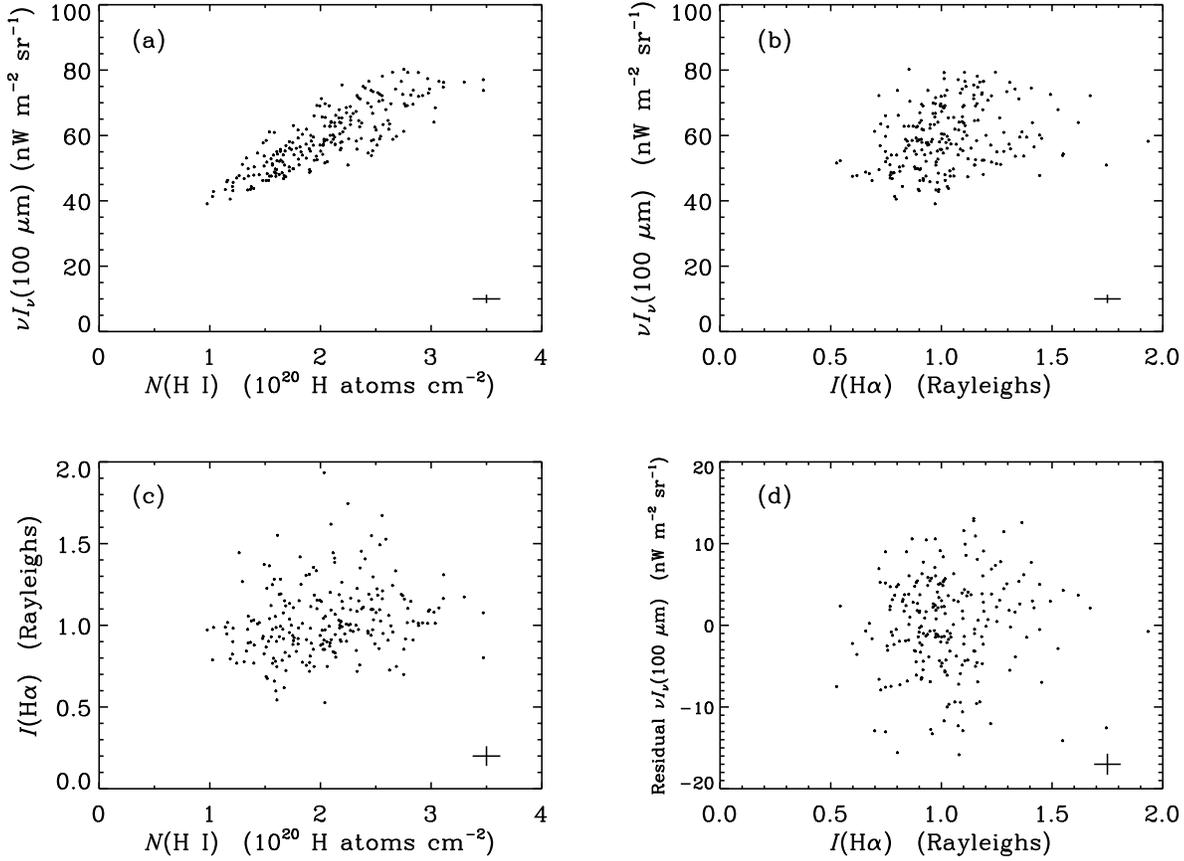}
\caption{Correlation plots for Q1 region positions used in our analysis, 
below an {\it N}(H~I) cut at 3 10$^{20}$ atoms cm$^{-2}$.  The correlation 
between 100 $\micron$ intensity and H~I column density (a) is tighter 
than that between 100 $\micron$ intensity and H$\alpha$ intensity (b) or that 
between H$\alpha$ intensity and H~I column density (c).  No correlation is seen
between H$\alpha$ intensity and residual 100 $\micron$ intensity after subtraction
of the H~I correlated component (d).  The cross plotted
in the lower right of each panel shows the typical statistical
measurement uncertainty ($\pm$ 1$\sigma$) for each quantity.}
\end{figure}
\clearpage

\begin{figure} 
\figurenum{4}   
\includegraphics[angle=90,width=\textwidth]{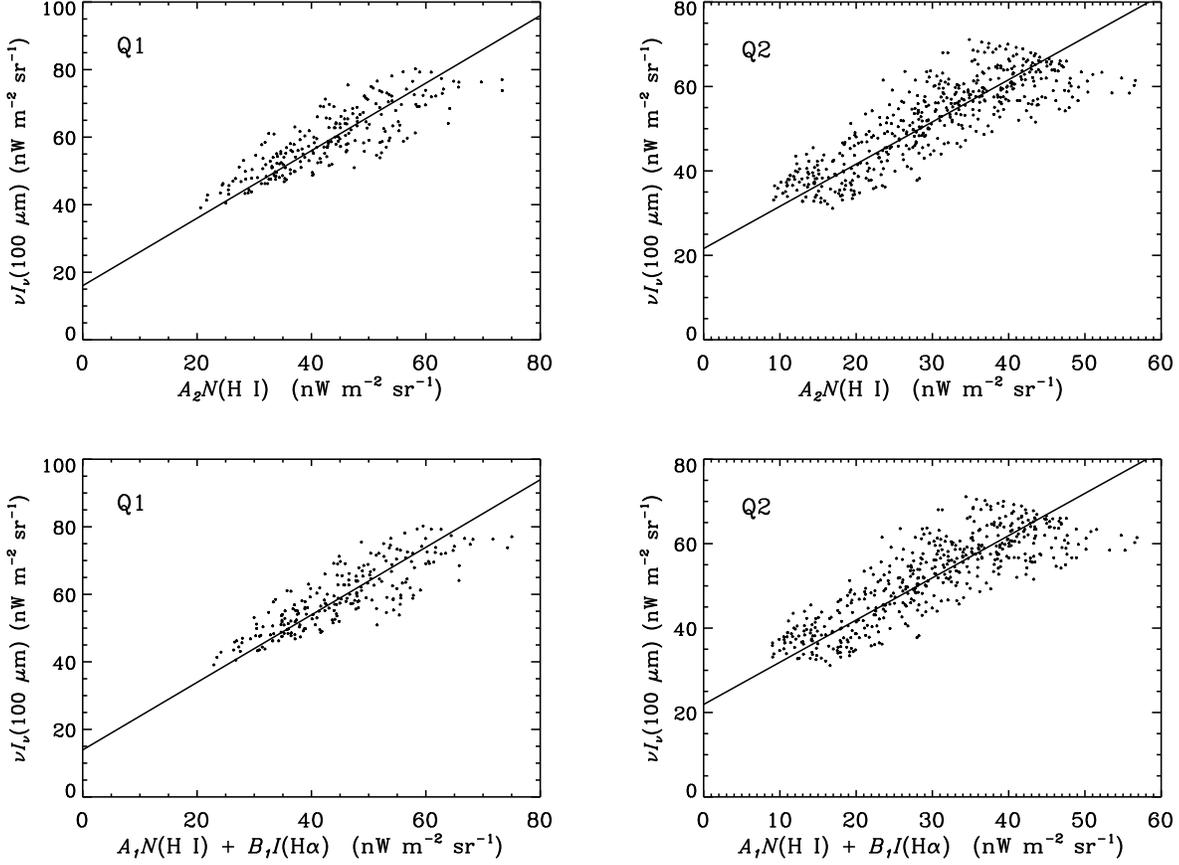}
\caption{Fits to the 100 $\micron$ data for the Q1 and Q2 regions.  For each
region, the 
100 $\micron$ intercept is nearly the same for the fit using H~I (top) and 
the fit using H~I and H$\alpha$ (bottom).  The scatter about the fit line is
also nearly the same for the two cases.}
\end{figure}
\clearpage

\begin{figure} 
\figurenum{5}                                                                                                                       
\epsscale{1.00}                                                                                                `
\plotone{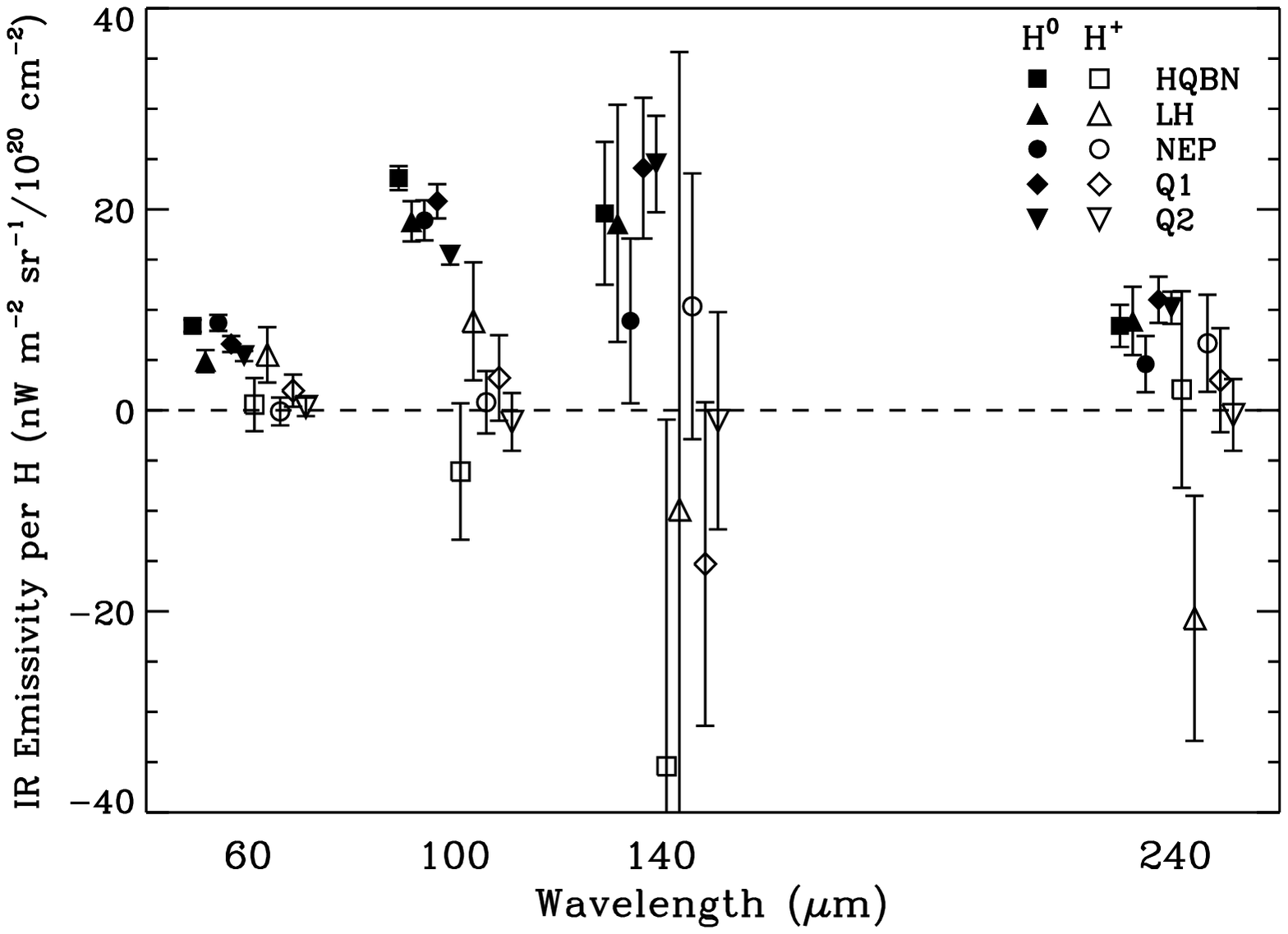}
\caption{Infrared emissivity per H nucleus for the neutral atomic gas
phase (filled symbols) and the ionized gas phase (open symbols) from the
three-component fits for each region.  For the ionized gas phase, the
emissivity values were obtained from the $B_1$ parameter values using a
conversion factor of $I(H\alpha$)/$N(H^{+})$ = 1.15 Rayleighs/$10^{20}$ 
cm$^{-2}$ (see text).  The emissivity values for the ionized phase are
consistent with zero, and the emissivity values for the neutral phase
are consistent with those derived when an H$\alpha$-correlated
component is not included in the fits (see Table 3).}
\end{figure}
\clearpage

\begin{figure} 
\figurenum{6}                                                                                                                       
\epsscale{1.00}                                                                                                `
\plotone{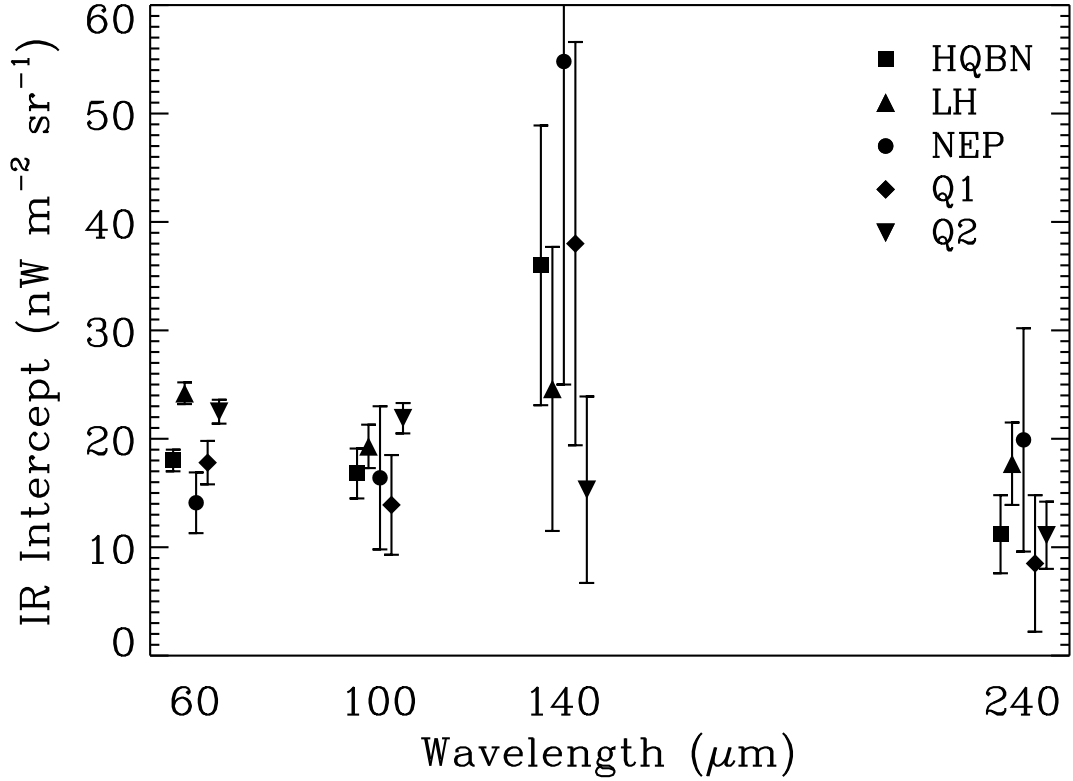}
\caption{Mean residual infrared intensity $C_1$ as a function of 
wavelength, from the three-component fits for each region.  The residual
intensity values for the five regions are consistent with isotropy
at 140 and 240 $\micron$, marginally consistent with isotropy at 100
$\micron$, and not consistent with isotropy at 60 $\micron$. The error 
bars show 1$\sigma$ statistical uncertainties and do not include systematic 
uncertainties that contribute to the total uncertainty for a CIB measurement.}
\end{figure}
\clearpage

\begin{figure} 
\figurenum{7}                                                                                                                       
\epsscale{1.00}                                                                                                `
\plotone{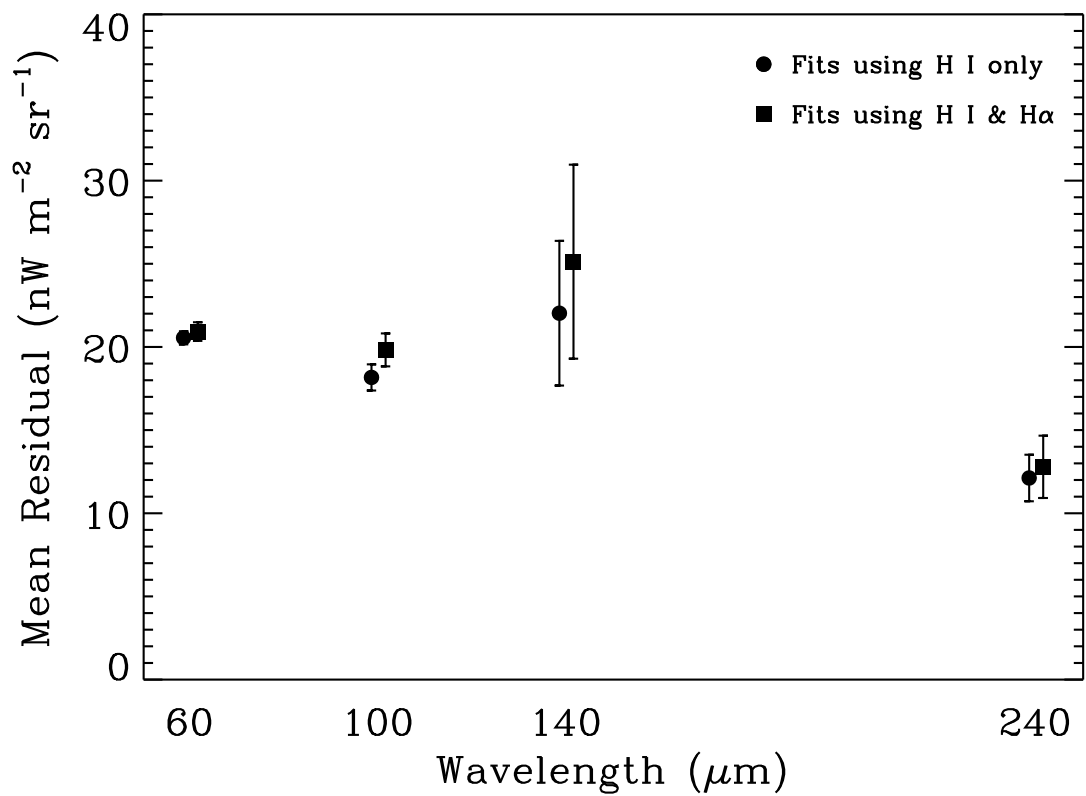}
\caption{Residual infrared intensity averaged over the HQBN, LH, NEP, 
Q1, and Q2 regions as a function of wavelength.  Results from the two-component 
fits (circles) and from the three-component fits (squares) are in close 
agreement.  The error bars show 1$\sigma$ statistical uncertainties and do not include 
systematic uncertainties that contribute to the total uncertainty for a CIB measurement.}
\end{figure}
\clearpage

\begin{figure} 
\figurenum{8}                                                                                                                       
\epsscale{1.00}                                                                                                `
\plotone{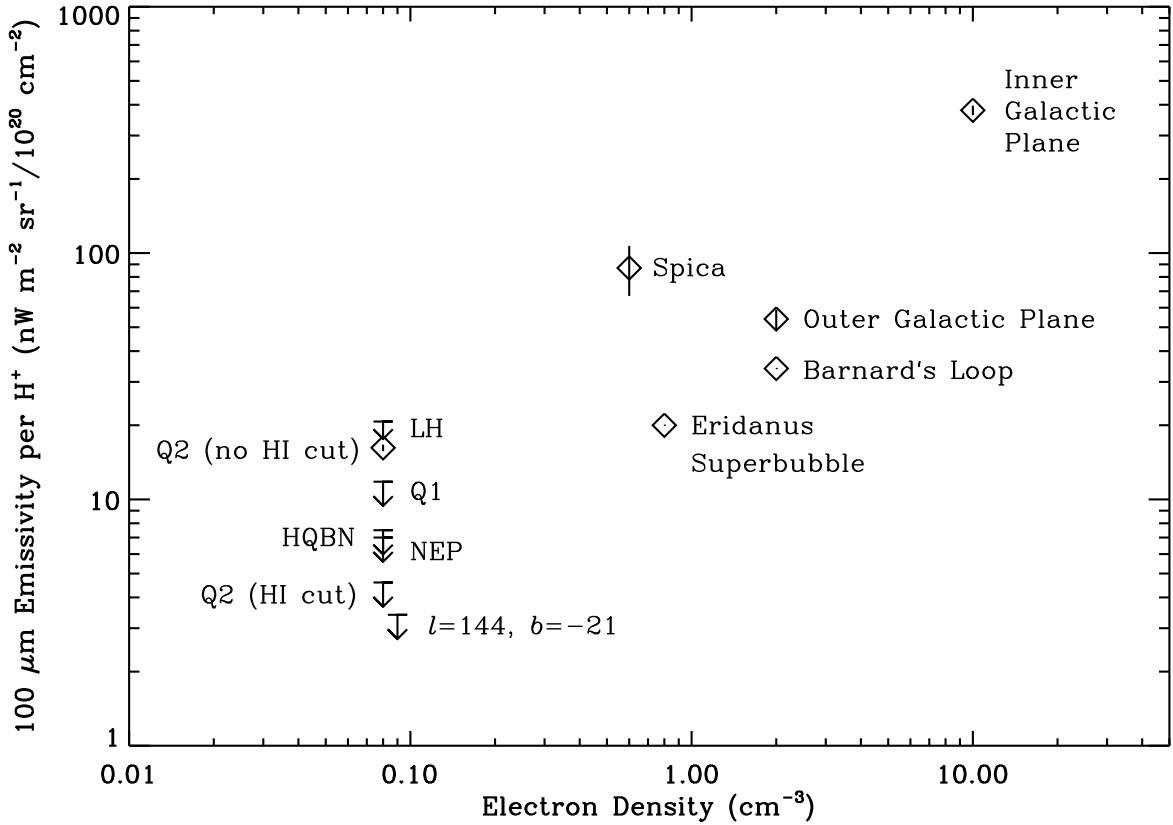}
\caption{Derived values of 100 $\micron$ emissivity per H$^+$ ion for
different regions plotted as a function of estimated electron density.  The data are
from Table 7.  Most previously studied regions have emissivities greater than
the 2$\sigma$ upper limits shown for our regions.  In most cases this can be 
attributed either to enhanced dust heating by nearby stars or, for regions with 
electron density of about 1 cm$^{-3}$ or greater, to enhanced dust heating by 
Ly$\alpha$ radiation.}
\end{figure}

\end{document}